\documentclass[11pt,a4paper]{article}
\pdfoutput=1

\usepackage{jcappub}
\usepackage{multirow}
\usepackage{amssymb} 
\usepackage[normalem]{ulem}

\newcommand{\fett}[1]{\boldsymbol{#1}}

\newcommand{\dd}{{\rm{d}}}
\newcommand{\ii}{{\rm{i}}}

\newcommand{\be}{\begin{equation}}
\newcommand{\ee}{\end{equation}}

\newcommand{\kop}
{\mathfrak{K}}

\definecolor{darkred}{rgb}{0.5,0,0}
\definecolor{darkgreen}{rgb}{0,0.6,0}
\definecolor{darkblue}{rgb}{0,0,0.5}
\definecolor{mypurple}{RGB}{180,20,180}

\newcommand{\Hc}{\mathcal{H}}

\newcommand{\CLASS}{{\sc class}}

\newcommand{\inspire}[1]{\href{http://inspirehep.net/search?p=find+J+#1}
 {{\color{black}[{\color{blue} {\small in}SPIRE}]}}}
\newcommand{\book}[1]{\href{http://inspirehep.net/search?p=#1}
 {{\color{black}[{\color{blue} {\small in}SPIRE}]}}}

\newcommand{\inspired}[1]{\href{http://inspirehep.net/search?p=#1}
 {{\color{black}[{\color{blue} {\small in}SPIRE}]}}}

\newcommand{\HL}{H_{\rm L}}
\newcommand{\HT}{H_{\rm T}}

\newcommand{\nab}{\nabla}

\makeatletter
\newsavebox\myboxA
\newsavebox\myboxB
\newlength\mylenA

\newcommand*\mybar[2][0.75]{%
    \sbox{\myboxA}{$\m@th#2$}%
    \setbox\myboxB\null
    \ht\myboxB=\ht\myboxA%
    \dp\myboxB=\dp\myboxA%
    \wd\myboxB=#1\wd\myboxA
    \sbox\myboxB{$\m@th\overline{\copy\myboxB}$}
    \setlength\mylenA{\the\wd\myboxA}
    \addtolength\mylenA{-\the\wd\myboxB}%
    \ifdim\wd\myboxB<\wd\myboxA%
       \rlap{\hskip 0.5\mylenA\usebox\myboxB}{\usebox\myboxA}%
    \else
        \hskip -0.5\mylenA\rlap{\usebox\myboxA}{\hskip 0.5\mylenA\usebox\myboxB}%
    \fi}
\makeatother

\begin{document}

\title{A new approach to cosmological structure formation with massive neutrinos}

\date{\today}

\author[a]{Christian Fidler,}
\emailAdd{fidler@physik.rwth-aachen.de}

\author[a]{Alexander Kleinjohann,}
\emailAdd{kleinjohann@physik.rwth-aachen.de}

\author[b]{Thomas Tram,}
\emailAdd{thomas.tram@phys.au.dk}

\author[c]{Cornelius Rampf,}
\emailAdd{rampf@thphys.uni-heidelberg.de}

\author[d]{Kazuya Koyama}
\emailAdd{kazuya.koyama@port.ac.uk}

\affiliation[a]{Institute for Theoretical Particle Physics and Cosmology (TTK), RWTH Aachen University, Otto-Blumenthal-Strasse, D--52057 Aachen, Germany.}

\affiliation[b]{Department of Physics and Astronomy, University of Aarhus, Ny Munkegade 120, DK--8000 Aarhus C, Denmark}

\affiliation[c]{Institut f\"ur Theoretische Physik, Universit\"at Heidelberg, Philosophenweg 16, D--69120 Heidelberg, Germany}

\affiliation[d]{Institute of Cosmology and Gravitation, University of Portsmouth, Portsmouth PO1 3FX, United Kingdom}

\abstract{
We show how Newtonian cosmological simulations can be employed to investigate the non-linear evolution of two particle species in a relativistic context. We discuss the application 
for massive neutrinos and other multi-species systems such as Cold Dark Matter (CDM) plus baryons or Warm Dark Matter (WDM). 
We propose a method that allows us to perform simulations including massive neutrinos and general relativistic effects at almost the same computational cost as ordinary CDM only N-body simulations, employing tailor-made initial conditions and a dictionary for the interpretation of the simulation output.
}

\maketitle   

\flushbottom
\section{Introduction}
\label{Introduction}

It is nowadays an established fact that the observed large-scale structure (LSS) is the consequence of non-linear amplifications of initially small fluctuations in the energy density. 
According to the $\Lambda$CDM paradigm, we live in a Universe that is composed of multiple species, i.e., cold dark matter (CDM),  baryons, neutrinos, photons and a cosmological constant ($\Lambda$). 
The gravitational dynamics of these species
is governed by the
Einstein--Boltzmann equations.

Solving the fully non-linear set of Einstein--Boltzmann equations is currently not feasible, thus one needs to resort to appropriate approximate solutions. For example, at sufficiently early times when energy fluctuations are still very small, the {\it linearised} Einstein--Boltzmann system delivers excellent approximations of the full system. Correspondingly, the approximation up to second-order of the Einstein--Boltzmann system yields only small corrections to the linear response \cite{Huang:2012ub,Su:2012gt,Pettinari:2013he}, indicating that cosmological perturbation theory (CPT; \cite{Kodama:1985bj,Malik:2008im,Villa:2015ppa}) may converge for sufficiently early times.

As time goes on, perturbations in the energy densities can become much larger than their initial values, especially on smaller scales. Therefore, one needs to resort to other solution techniques that go beyond CPT.
Common tools for that purpose are cosmological N-body simulations \cite{Teyssier:2001cp,Springel:2005mi,Hahn:2015sia}, where a large number of tracer particles are used to approximate the phase-space dynamics of matter to high accuracy. 
Apart from
a few exceptions \cite{Adamek:2015eda,Adamek:2016zes},
these simulations employ a Newtonian gravity solver.
Furthermore, since the CDM energy density is much larger than the baryon density, roughly by a factor of 5, it is customary to ignore the relative velocity differences and instead evolve a combined matter fluid (see however e.g.\ \cite{Angulo:2013qp,Valkenburg:2016xek}).

Another common simplification for such simulations is to ignore the evolution of massless radiation species -- although there is an increasing literature on how to incorporate these effects (see e.g.\ \cite{Fidler:2015npa,Fidler:2017ebh}).
Since these radiation perturbations remain sufficiently small over a wide range of cosmological times and scales, their evolution can be accurately described by linear perturbation theory.  By employing a linear Einstein--Boltzmann code, it is then possible to incorporate the radiation feedback on matter into Newtonian N-body simulations \cite{Brandbyge:2016raj,Adamek:2017grt}. This procedure has the added benefit of providing the General Relativistic (GR) metric of the simulation. 
Using the Newtonian-motion gauge approach (see section~\ref{sec:recap} for a recap), ref. \cite{Fidler:2016tir}, incorporated those linear relativistic effects as a kind of post-processing to standard Newtonian simulations.
Even more recently, this approach has been pushed to a much higher level of accuracy, namely by relaxing the {\it linear dictionary} between General Relativity (GR) and the Newtonian simulation, and instead employing a {\it weak-field dictionary} \cite{NLNM}.
In the present paper, building on top of the aforementioned works, we show how non-CDM species, massive neutrinos in particular, may be incorporated in a standard Newtonian N-body simulation.

With the ground-breaking discovery of flavour oscillations in neutrinos (see e.g.\ \cite{Tortola:2012te} for a review), it became clearly evident that neutrinos have a non-vanishing mass.
Massive neutrinos effectively behave as radiation at earlier times when the Universe is still hot, but eventually behave as matter.
This has very interesting consequences for the process of structure formation: massive neutrinos with large momenta have a significant pressure component which acts as a counter-force to gravity, thus prohibiting to some extent non-linear clustering, whereas neutrinos with low momenta collapse in a way similar to normal matter. Different methods have been suggested for including massive neutrinos directly in simulations. The two major methods for including neutrinos are either as a an additional particle species \cite{Brandbyge:2008rv,Brandbyge:2009ce,Viel:2010bn,VillaescusaNavarro:2012ag,Adamek:2017uiq,Banerjee:2018bxy} or as a fluid \cite{Viel:2010bn,Agarwal:2010mt,AliHaimoud:2012vj,Liu:2017now,Dakin:2017idt}. In the first case the challenge is including a sufficient number of neutrinos to represent their complex phase-space that has large thermal velocities at the time of initialisation. The description of neutrinos as a fluid avoids this problem, but faces its own challenges incorporating non-linear corrections that are important especially when the neutrinos have a large mass and different methods to simplify this problem are employed. On the other hand for lighter neutrinos the fluid approach is particularly efficient as the well understood linear evolution can be employed. Usually in these simulations relativistic contributions are neglected limiting the validity on the larger scales, with the exception of \cite{Adamek:2017uiq} where a full relativistic simulation of neutrinos is performed. There is no consensus yet on which method is the best, and to some degree   
that depends on the considered scales, as well as on the yet unknown neutrino masses. 

In the present paper, we provide a weak-field dictionary for Newtonian simulations that evolve at least two different species, plus radiation. 
We pay particular attention to the case where matter and massive neutrinos are considered,
but we also investigate other relevant cases such as CDM plus baryon simulations. For all these cases we show how to obtain a relativistic interpretation and determine the underlying weak-field space-time.

This paper is organised as follows.
In the next section we introduce our notations and conventions, as well as outline the employed weak-field expansion scheme.
In section~\ref{sec:recap} we review the basic Newtonian motion gauge approach. From there on, all our equations will be with respect to the temporal gauge choice of the well-known Poisson gauge (see e.g., \cite{Mukhanov:1990me,Ma:1995ey}; sometimes also called Newtonian gauge or longitudinal gauge). Before fixing the spatial gauge 
we turn to our novel relativistic description for
multiple species.
In particular, we outline the used assumptions
of our approach
in section~\ref{sec:multi}, 
provide the coupled evolution equations for a generic multi-species problem in section~\ref{sec:WFeq}, and fix the spatial gauge in section~\ref{sec:gauge}.
Then, in section~\ref{sec:aplications} we discuss a selection of specific multi-species problems, i.e., the cases of 
CDM and baryons  (section~\ref{sec:BCDM}),
CDM and a warm dark matter component (WDM, section~\ref{sec:CDMWDM}),
and finally the case of CDM and massive neutrinos (section~\ref{sec:nuCDM}).

\section{Definitions and weak-field expansion}

For simplicity, in this paper we only consider the evolution of scalar perturbations;
incorporating vector and tensor perturbations is however straightforward.
In our convention, the metric line element (summation over repeated indices is assumed, the speed of light is set to unity)
\be \dd s^2 = g_{\mu\nu} \,\dd x^\mu \dd x^\nu = g_{00}\, \dd \tau^2 + 2 g_{0i}\, \dd x^i \dd \tau + g_{ij} \,\dd x^i \dd x^j
\ee
in a yet unspecified gauge has the following metric coefficients of the scalar type:
\begin{subequations}
\label{metric-potentials}
\begin{align}
  g_{00} &= -a^2 \left[ 1 + 2 A \right] \,, \\
  g_{0i} &=  -a^2  \hat\nab_i  B  \,,\\
  g_{ij} &= a^2 \left[ \delta_{ij} \left( 1 + 2 \HL \right) + 2 \left(  \hat\nab_i \hat\nab_j + \frac {\delta_{ij}}  {3} \right) \HT  \right] \,.
\end{align}
\end{subequations}
We make use of the conformal time variable $\tau$, defined via $a\, \dd \tau = \dd t$, where $a=a(\tau)$ is the cosmological scale factor with evolution given by the usual Friedmann equations.
In order to make our weak-field expansion scheme as transparent as possible (see below), we make use of the normalised gradient operator 
$\hat\nab_i \equiv -(-\nabla^2)^{-1/2} \nab_i$ which translates into  $-\ii \hat{k}_i$ in Fourier space, where $\hat k_i \equiv k_i/|\fett{k}|$. 
This normalised gradient operator is known as the Riesz transform.
 
The employed weak-field scheme amounts to a double expansion scheme in metric perturbations and spatial derivatives, see \cite{NLNM} for a more thorough discussion.
Our weak-field expansion scheme is in agreement with the one of \cite{Brustein:2011dy,Kopp:2013tqa,NLNM}, whereas it differs in subleading terms from the schemes of  \cite{Green:2010qy,Green:2011wc,Adamek:2013wja}.
The weak-field approach  encompasses the largest cosmological scales where linear perturbation theory delivers an accurate description, but also incorporates the leading-order contributions from smaller scales where non-linear effects are important. 
We only demand the smallness of the metric perturbations and allow their source terms (the fluid variables) to become arbitrarily large.
We note that such a double expansion leads only to consistent equations for certain gauge choices; for example, a gauge that sets the fluid velocity equal to the metric perturbation $B$ trivially invalidates our choice of expansion. We choose a temporal gauge that preserves the weak-field limit, as discussed in \cite{NLNM}.

The evolution of the scalar perturbations $A$, $B$, $\HL$ and $\HT$ is given by the Einstein equations, which are sourced by the non-linear total stress-energy tensor, in space-time components 
\begin{subequations}
\label{energy-momentum}
\begin{align}
  T^0_{\phantom{0}0} &= -\sum_X \rho_X \equiv - \rho \,, \\
 T^0_{\phantom{0}i} &= \sum_X[\rho_X+ p_X] \hat\nab_i (  v_X -  B ) \equiv [\rho+ p] \hat\nab_i (  v -  B ) \,, \\ 
 T^i_{\phantom{i}j} &=  \sum_X \left[ p_X \delta^i_j  + \left(  \hat\nab^i \hat\nab_j + \frac {\delta^i_{j}} {3} \right) \Sigma_X  + [\rho_{X}+p_{X}] ( \hat\nab^i v_{X}) \hat\nab_j v_{X} \right] \nonumber \\ 
  & \equiv p \delta^i_j  + \left(  \hat\nab^i \hat\nab_j + \frac {\delta^i_{j}} {3} \right) \Sigma  + [\rho+p] (\hat\nab^i v) \hat\nab_j v \,,
\end{align}
\end{subequations}
where $X$ runs over the relevant species in the universe,
and $\rho_X$,  $p_X$, $\hat\nab_i v_X$ and $\Sigma_X$ are respectively the density, pressure, velocity and anisotropic stress of species $X$.
The corrections to the space-space part of the stress tensor proportional to the velocity squared result from its definition in the fluid rest frame. 
These corrections 
give rise to a mixing stress and pressure for the combined fluid in the first equation,
\begin{subequations} \label{eqs}
\begin{align}
  p  &= \sum \limits_X p_X + \frac{1}{3} \left[ \sum \limits_X (\rho_X + p_X )  (\hat\nab^i v_X) (\hat\nab_i v_X ) - (\rho +  p )  (\hat\nab^i v) (\hat\nab_i v ) \right] \,,\\
 \Sigma &= \sum \limits_X \Sigma_X + \frac{3}{2}\left( \hat\nab^i \hat\nab_j + \frac {\delta^i_{j}} {3} \right)\left[ \sum \limits_X  (\rho_X + p_X )  (\hat\nab^i v_X) (\hat\nab_j v_X ) - (\rho +  p )  (\hat\nab^i v) (\hat\nab_j v )\right] \,,
\end{align}
\end{subequations}
where, the terms in the squared brackets in both equations do generally not cancel exactly and thus lead to so-called mixing contributions. Using the space-time part of \eqref{energy-momentum}, these mixing contributions can be written as
\begin{subequations}
\label{mixing-def}
\begin{align}
  p_{\rm mix}  &= \frac{1}{3} \sum \limits_X (\rho_X + p_X )  (\hat\nab^i v_X) \hat\nab_i (v_X -  v )\,, \\
 \Sigma_{\rm mix} &= \frac{3}{2}\sum \limits_X \left( \hat\nab^i \hat\nab_j + \frac {\delta^i_{j}} {3} \right) (\rho_X + p_X )  (\hat\nab^i v_X) \hat\nab_j (v_X  - v ) \,,
\end{align}
\end{subequations}
where we have also included the numerical (and operational) prefactors in front of the square-bracketed terms in equations~\eqref{eqs}.
These mixing contributions vanish trivially for components that follow the mean velocity (i.e., for which we have $v_X = v$). 

\section{Newtonian motion gauge approach: Recap and temporal gauge choice} \label{sec:recap}

\subsection{Brief introduction to Newtonian motion gauges}
In a series of recent papers \cite{Fidler:2016tir,Fidler:2017ebh,NLNM}, we have developed the Newtonian motion (Nm) gauge framework.
The essential idea of this framework is to enable standard Newtonian simulations to obtain a relativistic description, that in particular includes the effect of radiation on structure formation.
Specifically, in \cite{Fidler:2016tir,Fidler:2017ebh} we used cosmological perturbation theory to track the evolution of the relativistic species and their feedback on the non-linear matter evolution, as well as provided the relativistic space-time to standard Newtonian simulations. These early works relied on a relativistic dictionary that was based on linear perturbation theory. Since (matter) fluid variables generally become large at late times and non-linear scales, this dictionary becomes a poor
approximation at such times and scales.
Subsequently,
the smallness assumption in the fluid variables has been relieved in \cite{NLNM}, where the Newtonian motion gauge framework in the weak-field limit has been employed.

For technical representations of the Newtonian motion framework we refer to the aforementioned works. 
Here we present its essential idea within a weak-field expansion. 
A consistent relativistic interpretation of Newtonian simulations requires that the gauge choice does not violate the underlying smallness assumptions. 
In such gauges, the impact of GR introduces several non-Newtonian terms in the equations of motion for matter. In particular, the relativistic Euler equation that dictates momentum conservation generally includes GR corrections that can be computed by using Einstein--Boltzmann solvers.
At first sight, one way to incorporate these GR corrections would be to modify existing Newtonian simulations in such a way that the relativistic fluid equations are approximately satisfied
(for successful implementations see \cite{Brandbyge:2016raj,Adamek:2017grt}).

A different, almost orthogonal approach, is the Newtonian motion framework, that solves the task without the need to modify existing Newtonian simulations. Instead, the power of gauge freedom is used to employ a spatial gauge condition that precisely absorbs all GR corrections in the relativistic Euler equation of matter by definition. 
In such a Newtonian motion gauge, the relativistic motion thus matches the one of Newtonian gravity,
whereas the corresponding space-time is obtained by the Newtonian motion framework. Here it is important to note that the Newtonian simulation does not need to know about the dynamically changing space-time; instead the relativistic space-time is obtained as a post-processing that can be 
obtained after the Newtonian simulation is completed.

\subsection{Temporal gauge choice}\label{ssec:temporalgauge}
There is no unique Newtonian motion gauge since
for any fixed temporal gauge there exists a gauge where the relativistic Euler equation matches the Newtonian one for matter. There are however some temporal gauge choices which result in a formal violation of the underlying perturbative weak-field expansion, 
since a weak-field expansion demands the smallness of metric and gravitational potentials.

In \cite{NLNM} a  convenient temporal gauge choice was found to be the one of the Poisson gauge, as it implies, even in the presence of relativistic species, perturbative smallness of the metric/gravitational potentials in the weak field limit.
We thus stick to the same temporal gauge choice for the whole present paper given by 
\be 
 \label{temporalgauge}
\kop B =  \dot{H}_{\rm T} \,,
\ee
where we have introduced the operator $\kop = (-\nabla^2)^{1/2}$ that corresponds to the magnitude $k = |\fett{k}|$ in Fourier space.
The spatial gauge condition, chosen within the Newtonian motion class, will be defined in section~\ref{sec:gauge}. This means that we will define a gauge in which the same spatial hyper-surfaces as in the Poisson gauge are selected, but within these the coordinates are evolving such that particles follow a Newtonian motion within these coordinates. 

\subsection{The N-boisson gauge} \label{sec:Nboisson}

As evident from previous works (e.g.\ \cite{NLNM}), we
have seen that in a pure cold dark matter cosmology there exists a simple solution to the Newtonian motion gauge equations, where the spatial gauge is entirely fixed by the comoving curvature perturbation $\zeta$ via $\HT = 3\zeta = \rm{const}$.\footnote{Within our metric conventions, the comoving curvature perturbation is
$\zeta = \HL + \HT/3 - {\cal H} \kop^{-1}(v -B)$.}
This gauge is the weak-field gauge in which ordinary pure CDM Newtonian simulations can be brought into agreement with General Relativity and which will serve as a point of reference for the more complex situations we discuss in this paper. In \cite{Fidler:2015npa}, the N-body gauge was  introduced as a gauge with the spatial gauge condition $\HT= 3 \zeta$ on the temporal slicing of the comoving-orthogonal gauge. As discussed in section \ref{ssec:temporalgauge}, the temporal slicing of the Poisson gauge is more suitable for a weak-field analysis. Therefore we define the gauge combining the spatial coordinates of the \textbf{N-bo}dy gauge with the temporal slicing of the Po\textbf{isson} gauge; the N-boisson gauge ($\HT = 3\zeta$, $\kop B =  3\dot{\zeta}$).
We note that the N-boisson gauge is highly related to the analysis of \cite{Chisari:2011iq}, where the temporal {\it and} spatial gauge condition of the Poisson gauge is used, but in their dictionary the N-body particles are displaced to the spatial position of the N-body gauge essentially implementing the N-boisson gauge.

In the N-boisson gauge, the densities are identical to the Poisson gauge densities since they only depend on the shared temporal gauge choice. Velocities on the other hand are transformed by the time-derivative of the spatial gauge transformation and since $\zeta$ becomes constant in a matter dominated Universe these converge to the Poisson gauge values. In that case the N-boisson gauge differs from the Poisson gauge only in the higher moments that are affected by the introduction of the tensor metric perturbation $\HT \neq 0$.

\section{Multi-species Newtonian motion gauges}

\subsection{Strategy and assumptions}\label{sec:multi}

In this work we study multi-species systems such as the combination of baryons and cold dark matter. In general, the gauge freedom of Newtonian motion gauges is sufficient to set the dynamics of one species to be Newtonian, but in the case of two (or more) independent species it is not possible to have all species follow Newtonian dynamics. 

We resolve this tension by defining a combined fluid that tracks the centre-of-mass motion of the massive parts of all species that we want to simulate. We then fix a gauge such that this fluid evolves according to Newtonian dynamics and the N-body particles represent this combined massive fluid. This method uniquely describes the dynamics, since the metric potentials are only sensitive to the total energy-momentum tensor.
However, we do not obtain the full dynamics of the system as we do not solve for the relative motions of the individual species with respect to the shared fluid. We refer to appendix \ref{sec:resolve} for different strategies to resolve this secondary problem.

We consider an arbitrary amount of species $X$ and decompose them into a relativistic-like and a massive-like component, defined via
\be
\rho_m = \sum \limits_{X} \rho_X - 3 p_X\,, \quad \qquad
\rho_\gamma = \sum \limits_{X} 3p_X \,,
\ee
from which $\rho = \rho_m + \rho_\gamma$ and $\rho + p = \rho_m + \frac 4 3 \rho_\gamma$ follows trivially. Our definition is such that a cold massive species (like CDM) will only contribute to $\rho_m$, while a fully relativistic species (like photons) only contributes to $\rho_\gamma$. Warm components contribute to both $\rho_m$ and $\rho_\gamma$, and this decomposition is in general time-dependent.

We split the velocities according to 
\be
(\rho + p) v^i = \rho_m v^i_m + \frac 4 3 \rho_\gamma v^i_\gamma\,,
\ee
which means that $v_m$ is the center-of-mass velocity of the massive fluid $\rho_m$: $\rho_m v_m^i = \sum \limits_{X} (\rho_X -3 p_X) v_X^i$ and $v_\gamma$ is the velocity of the relativistic components $\rho_\gamma v_\gamma^i = \sum \limits_{X} 3 p_X v_X^i$. In the simple two species case of CDM and photons we have 
$\rho_m = \rho_\text{cdm}$, $v_m = v_\text{cdm}$, $\rho_\gamma = \rho_\text{photons}$, $v_\gamma = v_\text{photons}$.

As pointed out 
in the paragraph around equation~\eqref{mixing-def}, the total anisotropic stress and pressure receive mixing contributions proportional to the velocities squared when the species do not follow the mean velocity. For relativistic species these contributions are perturbatively suppressed and may be neglected, and in this case the total anisotropic stress is simply the sum of all individual stresses. The same is true for the mixing between relativistic and massive species. But the mixing between two massive species may be important and cannot be neglected in the weak-field approximation. We therefore 
write
\be
\Sigma = \Sigma_m + \Sigma_\gamma \,, \quad \qquad \delta p = \delta p_m + \delta p_\gamma \,,
\ee
where $\Sigma_\gamma$ and $\delta p_\gamma$ are the relativistic anisotropic stresses and pressures, and $\Sigma_m$, $\delta p_m$ are given by the massive stresses and pressures including $\Sigma_{\rm{mix}}$ and $\delta p_{\rm{mix}}$, quadratic in the velocities of the massive species.  

Under this assumption we see that the energy-momentum tensor can be built from the massive fluid $m$ and the relativistic fluid
 $\gamma$ by simple superposition. Our goal will therefore be the simulation of these two fluids where the process of disentangling into the individual species is a secondary problem that we discuss in Appendix \ref{sec:resolve}. Our strategy is to evolve the massive component in a non-linear particle simulation, while the relativistic part is solved by using a modified Einstein-Boltzmann code. The latter code may be based on linear perturbation theory which is justified since in the case of the relativistic species, its pressure prevents the collapse, thus its perturbations remain sufficiently small at late times.
The N-body particles then no longer represent one single species, but track the center-of-mass evolution of the massive-like species.

\subsection{Weak-field evolution equations}\label{sec:WFeq}

When we translate the above summarised weak-field assumptions for the individual species
into assumptions for the combined fluids $m$ and $\gamma$, it follows that the relativistic component always remains of order $\epsilon$ (the amplitude of initial fluctuations) as it is prevented from growth (we ignore higher-order corrections in $\epsilon$ as they remain vanishingly small for the relativistic component, even at late times). In our expansion scheme, the leading-order terms for the massive density are of order $\kappa^2 \epsilon$, the velocity of order $\kappa \epsilon$ and all higher moments are of order $\epsilon$, including the mixing stress, where $\kappa$ is the expansion parameter related to spatial gradients. 

In addition we make the assumption that all relativistic perturbations can, at all times, be extracted from a linear Einstein-Boltzmann code and any perturbation $X_\gamma$ satisfies $X_\gamma = X_\gamma^{(1)}$. Note that this assumption is not fundamental to the following discussion, but it allows an efficient computation of the relativistic perturbations from a linear Einstein--Boltzmann code. 

We first derive the non-linear weak-field equations for the massive components. We start from the weak-field Euler equation, obtained from momentum conservation
\begin{align} \nonumber
\left(\partial_\tau +4\Hc\right)& (\rho+p) \hat\nab_i v - (\rho+p) (\hat\nab_i \kop A + \left(\partial_\tau +\Hc\right) \kop^{-1} \hat\nab_i  \dot{H}_{\rm T})) \\ 
&=  \dot{p} \kop^{-1} \hat\nab_i \dot{H}_{\rm T} - 3(\rho+p) \dot{H}_{\rm L}\hat\nab_i v + \kop\hat\nab^j \left[(\rho + p)(\hat\nab_i v)(\hat\nab_j v) \right] +  \kop\hat\nab_i (\delta p - \frac 2 3 \Sigma)  \,,
\end{align}
while the continuity equation reads
\be
\dot{\rho} = -3\Hc (\rho+p) - 3(\rho+p)\dot{H}_{\rm L} + \kop\hat\nab^i \left[(\rho+p)\hat\nab_i v\right].
\ee
We split these equations into the massive and relativistic fluid. 
This task is straightforward since only combinations of the energy-momentum tensor appear in these equations.
Note however, that the combined stresses and pressures are not just the sum of the individual ones, and there are non-linear mixing corrections (see equations~\eqref{eqs}).
Combining the last two equations we find
\begin{align} \nonumber
\rho_m \left(\partial_\tau +\Hc\right)&\hat\nab_i v_m - \dot{\bar{p}}( \kop^{-1} \hat\nab_i \dot{H}_{\rm T} + 3 \hat\nab_i v_m ) - 4 \Hc \rho_\gamma \hat\nab_i v_m  \\ \nonumber
& - (\rho_m + \frac 4 3 \rho_\gamma) (\hat\nab_i \kop A + \left(\partial_\tau +\Hc\right) \kop^{-1} \hat\nab_i  \dot{H}_{\rm T}) \\
 =&   \rho_m \hat\nab_j v_m \kop\hat\nab^j\hat\nab_i v_m  +  \kop\hat\nab_i (\delta p_m + \delta p_\gamma - \frac 2 3 \Sigma_m - \frac 2 3 \Sigma_\gamma) -  C_{i,\gamma}   \,,
\end{align}
where 
$ C_{i,\gamma} = (4/3) \left(\partial_\tau +4\Hc\right)  \rho_\gamma  \hat\nab_i v_\gamma $ 
which describes the evolution of the relativistic components that may be computed to linear order in perturbation theory. Replacing $\rho_{\gamma}\hat\nab_i v_{\gamma}$ by a sum over all species and using the Euler equation for each of them we find:
\begin{align}
C_{i,\gamma} =& 4 \Hc \rho_\gamma \hat\nab_i v_\gamma  + \frac 4 3 \rho_\gamma (\hat\nab_i \kop A + \left(\partial_\tau +\Hc\right) \kop^{-1} \hat\nab_i  \dot{H}_{\rm T}) \nonumber \\
&+ \sum \limits_{X} 4 \dot{p}_X \hat\nab_i v_X  - \frac{4 p_X}{ \rho_X + p_X} \left( \dot{p}_X(\hat\nab_i v_X - \kop^{-1}\hat\nab_i \dot{H}_{\rm T}) -\hat\nab_i \kop \delta p_X + \frac 2 3 \hat\nab_i \kop \Sigma_X \right)\,.
\end{align}
In total this provides the expression
\begin{align} \nonumber
(\partial_\tau + & \Hc)\hat\nab_i v_m -\hat\nab_i \kop A - \left(\partial_\tau +\Hc\right) \kop^{-1} \hat\nab_i  \dot{H}_{\rm T} - \hat\nab_j v_m \kop\hat\nab^j\hat\nab_i v_m - \frac {\kop\hat\nab_i} {\rho_m} (\delta p_m - \frac 2 3 \Sigma_m)  \\
=& - 4 \Hc \frac{\rho_\gamma}{\rho_m} \hat\nab_i (v_\gamma - v_m) + \frac{\dot{\bar{p}}}{\rho_m} ( \kop^{-1} \hat\nab_i \dot{H}_{\rm T} + 3 \hat\nab_i v_m ) + \frac {\kop\hat\nab_i} {\rho_m} ( \delta p_\gamma  - \frac 2 3 \Sigma_\gamma) \nonumber \\
& - \sum \limits_X \left[ 4 \frac{\dot{p}_X}{\rho_m} \hat\nab_i v_X  - \frac{4 p_X}{ (\rho_X + p_X)\rho_m} \left( \dot{p}_X(\hat\nab_i v_X - \kop^{-1}\hat\nab_i \dot{H}_{\rm T}) -\hat\nab_i \kop \delta p_X + \frac 2 3 \hat\nab_i \kop \Sigma_X \right)\right]\,,
\end{align}
where the first line is the non-linear Vlasov-Poisson equation for the massive species and the following lines are corrections from the presence of relativistic components and metric perturbations. 

The continuity equation separated into massive and relativistic components yields
\begin{align}
\dot{\rho}_m =& -3\Hc\rho_m - 3\rho_m\dot{H}_{\rm L} + \kop\hat\nab^i\rho_m\hat\nab_i v_m \nonumber \\  
&- \dot{\rho}_\gamma  -4\Hc\rho_\gamma - 4\rho_\gamma\dot{H}_{\rm L} + \frac 4 3 \rho_\gamma \kop\hat\nab^i\hat\nab_i v_\gamma\,,
\end{align}
where again the first line corresponds to the non-linear Vlasov-Poisson equation for the massive fluid $\rho_m$, while the lower line consists of corrections from the metric and the relativistic species.

The equations of motion for the higher massive kinetic moments are given by the Newtonian Vlasov-Poisson equations. Relativistic sources are only found in the lowest multipoles since the metric is only a rank two tensor and cannot directly couple to higher moments to leading weak-field order. The same is found in the linear Boltzmann equations for the relativistic species. Higher moments are sources from the lower ones by multi-stream mixing and indirectly influenced by gravity via its impact on the lowest moments. 

For the relativistic species we employ the well-known linear Einstein--Boltzmann equations and combine the fluids weighted by their respective pressures. This provides a hierarchy of equations for all moments.

\subsection{The spatial gauge condition}\label{sec:gauge}

We have seen that we can decompose the relativistic weak-field equations of motion into a modified non-linear Vlasov-Poisson equation for a combined massive fluid plus Einstein-Boltzmann equations for the relativistic components. We represent the massive fluid by particles and evolve them in a non-linear N-body simulation, reproducing the Vlasov-Poisson dynamics. Our goal is thus to absorb the non-Newtonian corrections appearing in the lowest moments of the Vlasov-Poisson equations.

We start with the continuity equation where we find two types of corrections that are not present in a Newtonian simulation. First we have the volume perturbations, stemming from a non-vanishing perturbation in the trace part of the spatial metric \mbox{(cf.\,\cite{Fidler:2015npa,Adamek:2017grt}).} Secondly, we
find corrections induced from the relativistic mass modulation:
If we have a component that is initially relativistic, but massive at the end of the simulation, it will be counted as $\rho_\gamma$ initially, but $\rho_m$ towards the end. This means that during the transition its perturbations are imprinted on the massive fluid. 
To absorb these we define the simulation density
\be
\rho_N = \rho_m + 3\HL \rho_m - M \rho_m \,,
\label{definition of simulation density}
\ee
where the factor $(1+3\HL)$ incorporates said volume perturbation, and the mass modulation $M \equiv \bar{M} + \delta M$
is required to track the conversion from relativistic into massive species. 
In cosmologies without such a conversion/modulation, it vanishes. Since the corrections to the continuity equation are small in the weak-field sense we count $M = \mathcal{O}(\epsilon)$.
From requiring that the Newtonian density follows the unmodified Vlasov-Poisson equation
\be
\dot{\rho}_N = -3\Hc\rho_N + \kop\hat\nab^i\rho_N\hat\nab_i v_m \,, 
\ee
we obtain the equations for the background mass modulation
\be
\dot{\bar{M}} = (1 - \bar{M})(3\Hc + \frac{\dot{\bar{\rho}}_m}{\bar\rho_m}) \,,
\ee
and for its perturbation
\begin{align} \nonumber
\delta \dot{M} + (3\Hc + \frac{\dot{\bar{\rho}}_m}{\bar\rho_m}) \delta M = \;
&3(3\Hc +\frac{\dot{\bar{\rho}}_m}{\bar\rho_m}) \HL 
- \dot{\bar{M}} \delta_m
+ 3 \bar{M} \dot{H}_{\rm L} \\
- &\frac{1 - \bar{M}}{\bar\rho_m}\left[\delta\dot{\rho}_\gamma + 4\Hc\delta\rho_\gamma + 4\Hc\bar{\rho}_\gamma\dot{H}_{\rm L} - \frac 4 3 \bar{\rho}_\gamma \kop\hat\nab^i\hat\nab_i v_\gamma\right] \,.
\end{align}
Physically the mass modulation can be understood as a post-processing change of the masses of the N-body particles to incorporate that a previously relativistic species has become massive and is now included in $\rho_m$. The initial values for $\bar{M}$ and $\delta M$ are free parameters and could be set to zero, meaning that we initialise the system using the initial matter distribution. Alternatively we may also set the present day mass modulation to zero, meaning that we will have the correct amount of mass in the simulation at the late times when non-linear effects are relevant. This seems advantageous as the late evolution is highly non-linear. The other sources only depend on the relativistic fluid and can be accurately computed in linear theory.

The final step is fixing the remaining spatial gauge freedom such that the Euler equation takes the Newtonian form
\be 
\left(\partial_\tau +\Hc\right)\hat\nab_i v_m + \hat\nab_i \kop \Phi_N  - \hat\nab_j v_m \kop\hat\nab^j\hat\nab_i v_m - \frac {\kop\hat\nab_i} {\rho_m} (\delta p_m - \frac 2 3 \Sigma_m) = 0 \,,
\ee
with
\be
\kop^2 \Phi_N =  4\pi G a^2 \delta\rho_N \,,
\ee
i.e., the Newtonian potential that is derived from the Newtonian density alone.
It then follows directly that the necessary Newtonian motion gauge condition is  
\be
\left(\partial_\tau +\Hc\right) \kop^{-1} \hat\nab_i  \dot{H}_{\rm T} + \hat\nab_i \kop A + T_\gamma = - \hat\nab_i \kop \Phi_N\,,
\label{spgauge}
\ee
where $T_\gamma$ are the corrections due to the relativistic components and reads
\begin{align} \nonumber
T_\gamma=& - 4 \Hc \frac{\rho_\gamma}{\rho_m} \hat\nab_i (v_\gamma - v_m) + \frac{\dot{\bar{p}}}{\rho_m} ( \kop^{-1} \hat\nab_i \dot{H}_{\rm T} + 3 \hat\nab_i v_m ) + \frac {\kop\hat\nab_i} {\rho_m} ( \delta p_\gamma  - \frac 2 3 \Sigma_\gamma)\\
& - \sum \limits_X \left[ 4 \frac{\dot{p}_X}{\rho_m} \hat\nab_i v_X  - \frac{4 p_X}{ (\rho_X + p_X)\rho_m} \left( \dot{p}_X(\hat\nab_i v_X - \kop^{-1}\hat\nab_i \dot{H}_{\rm T}) -\hat\nab_i \kop \delta p_X + \frac 2 3 \hat\nab_i \kop \Sigma_X \right)\right] \,.
\label{T gamma}
\end{align}
These are related to the relativistic fluids only and thus are perturbatively small and accurately computed in linear theory. When there is no energy exchange between the combined massive fluid and the combined relativistic fluid, all corrections cancel and we get $T_\gamma=0$. As in the single fluid case the difference of $\Phi_N$ and $A$ appears and is given by  
\be
\kop^{2} (A +\Phi_N)  = - 4\pi G a^2 \left[ \delta\rho_\gamma +\rho_m (\HT - 3\zeta + M) + 4 \Hc  \rho_\gamma \kop^{-1}(v - \kop^{-1}\dot{H}_{\rm T}) + 2 \Sigma \right]  \,,
\label{Aphi}
\ee
with the curvature perturbation 
\be
 \zeta = \HL + \frac 1 3  \HT - \Hc \kop^{-1}( v -  \kop^{-1}\dot{H}_{\rm T}) \,.
\ee
The mass modulation enters in the difference of the potentials, describing that mass is missing in the particle simulation, leading to incorrect forces. 
Using (\ref{spgauge}) and (\ref{Aphi}), the gauge condition now reads
\begin{align} \nonumber
\left(\partial_\tau +\Hc\right)&\dot{H}_{\rm T} \\ 
=& 4\pi G a^2 (\delta\rho_\gamma +\rho_m (\HT - 3\zeta + \delta M + \bar{M} \delta_m) + 4 \Hc  \rho_\gamma \kop^{-1}(v - \kop^{-1}\dot{H}_{\rm T}) + 2 \Sigma) - \kop T_\gamma \,,
\end{align}
and all sources are small in the weak-field sense. In addition, as in the single fluid case, the small scale impact of the gauge condition is vanishing as long as $\HT = \mathcal{O}(\epsilon)$ which implies that we may solve it using only the linear equations.

We have now constructed a gauge in which the variables $\rho_N$, $v_N = v_m$ and $\Phi_N$ obey the Newtonian  Vlasov-Poisson equation. We may solve this set of equations by representing them with N-body particles that evolve under Newtonian gravity. This system of equations is supplemented by relativistic but linear equations for the relativistic variables $\rho_\gamma, \, v_\gamma$, the metric potentials and the mass modulation $M$, which can be solved in a linear Einstein--Boltzmann code.

In the case of only CDM and photons it is trivial to see that $M=0$. Thus, in that case we recover the equations found for the single fluid limit in \cite{NLNM}.

\section{Application to a Newtonian simulation}\label{sec:aplications}

In the following we outline the applications of 
our novel framework. 
We will discuss a few special cases to illustrate the method. 

\subsection{CDM and baryons}\label{sec:BCDM}

First, we discuss the case of baryons and CDM as an example of a mixture of two massive species.

Since both fluids represent cold massive species we may employ the simple N-boisson gauge [cf. section~\ref{sec:Nboisson}] (or a suitable single fluid Nm gauge when considering radiation) to obtain Newtonian evolution equations for both species at once. This however requires that we initialise the simulation with two distinct species, each having their own velocity and density distributions. 

In practice in most simulations this is not done and instead a combined fluid with a single valued density and velocity is employed.
We now utilise our formalism of shared Nm gauges to examine the suitability of this approach.

It follows trivially that in this case $M=0$ since the mass is conserved.
The multi-fluid gauge condition is then
\begin{align} \nonumber
\left(\partial_\tau +\Hc\right)&\dot{H}_{\rm T} \\ 
=& 4\pi G a^2 (\delta\rho_\gamma +\rho_m (\HT - 3\zeta) + 4 \Hc  \rho_\gamma \kop^{-1}(v - \kop^{-1}\dot{H}_{\rm T}) + 2 \Sigma) \,,
\end{align}
where the index $\gamma$ denotes photons and massless neutrinos and we recover the N-boisson gauge in the simplest case of neglecting radiation perturbations. This is not surprising as it is already a Newtonian motion gauge for both species individually. On the level of the metric, both approaches of either employing a shared fluid or two separate fluids, are therefore equivalent. 
The initial conditions in both cases, however, are not the same. A simulation with two distinct fluids automatically incorporates the full mixing pressure and stress. In the shared case, we should instead incorporate the mixing stress into the initial conditions of our single massive fluid. 

A quick analysis reveals that this anisotropic stress is suppressed at early times and depends on the velocity difference, a source free quantity that decays over the lifetime of the Universe. Therefore it is a very good approximation to neglect this stress and initialise the simulation using the combined centre-of-mass variables $\rho_m$ and $v_m$.\footnote{Note that this does not imply that we ignore other sources of anisotropic stresses such as late time shell crossing.} This is an assumption that is commonly employed in N-body simulations and which we argue here from the relativistic point of view.

\paragraph{Concentration of the individual fluids.}

In Appendix \ref{sec:resolve} we derive a method to distinguish between the fluids and calculate their individual distributions. 
Here we summarise the results and refer the interested reader to the Appendix for a detailed derivation.
We define the local concentration coefficient:
\be
\sigma = \frac{\delta_{\rm cdm} - \delta_{b}}{1+\delta_N} \,.
\ee
Using this the baryon fraction is then:
\be
\frac{\bar{\rho}_{\rm b}}{\bar\rho} (1 + \frac{\bar{\rho}_{\rm cdm}}{\bar\rho}\sigma)\,.
\label{Definition of Baryon Fraction}
\ee

\begin{figure}
\includegraphics[width=.92\textwidth]{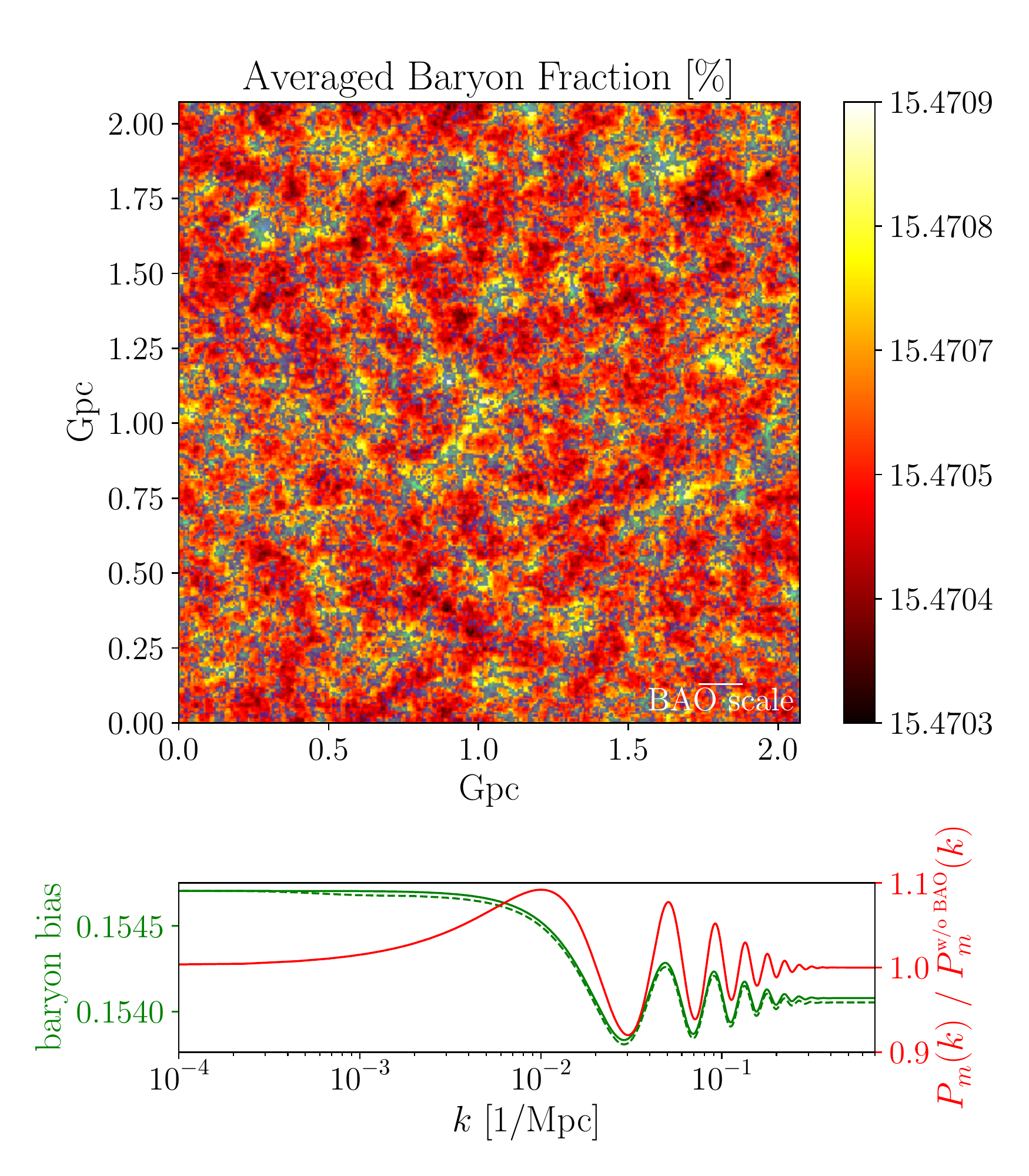} 
\caption{The upper plot shows a real space realisation of the baryon fraction $\bar\rho_b/\bar \rho_m$ defined in Eq.\,(\ref{Definition of Baryon Fraction}) averaged over $6.96$ Mpc in the third dimension. 
In blue, we have plotted the high density regions from an N-body simulation with the same initial seeds. The density has been integrated over the same region in the third dimension. The gevolution code \cite{Adamek:2016zes} was used for the simulation. The lower plot shows a linear estimate of the baryon bias $b = \delta\rho_{m,tot} / \delta\rho_b$ and the linear matter power spectrum relative to a smoothed linear matter power spectrum without baryonic acoustic oscillations (BAOs) for comparison. The solid line is in a cosmology without reionisation, while the dashed line includes reionisation. We calculated the smoothed spectrum by removing the BAO feature in its Discrete Sine Transform (DST) as described in Appendix A1 of ref.~\cite{BAOSmoothingMethod}. }
\label{Baryon Bias Plot}
\end{figure}

A typical baryon distribution and the high density regions of the corresponding N-body simulation can be seen in figure~\ref{Baryon Bias Plot}. It is to be noted that in regions with a higher total matter density, there appears to be a higher percentage of baryons. The scale of these regions is comparable to the BAO scale. The linear baryon bias plotted below also shows oscillatory features which coincide with the BAOs in the matter power spectrum plotted for comparison. This suggests that high density areas formed close to the BAO peaks consequently have a slightly higher than average baryon fraction.

Observationally this is unlikely to have an impact since the computed changes in the ratio between baryons and CDM remain very small. 

\paragraph{Impact of reionisation.}

Another interesting aspect is the impact of baryonic physics on N-body simulations. A feature that can easily be accounted for in our method is the chance for baryons to scatter off CMB photons during reionisation. The linear Einstein-Boltzmann code \CLASS~already takes this into account when computing the baryon and photon velocities, so we just need to add the collisional source
\be
\left(\partial_\tau +\Hc\right)\hat\nab_i v_b \mathrel{{+}{=}} \mathcal{R} \sigma_{\rm T} \hat\nab_i (v_\gamma-v_b) \,,
\ee
with
\be
\mathcal{R} = \frac{4\bar{\rho}_\gamma}{3\bar{\rho}_b} \,,
\ee
and the Thomson scattering rate $\sigma_{\rm T}$ to our gauge condition. Since this only affects the baryons, the term is suppressed by a factor $\rho_b/\rho_m$ in the final formula. It then acts as a source in the differential equation for $\HT$ which is relevant only during reionisation and is quickly diluted from the expansion of the Universe afterwards. 

\begin{figure}
\includegraphics[width=\textwidth]{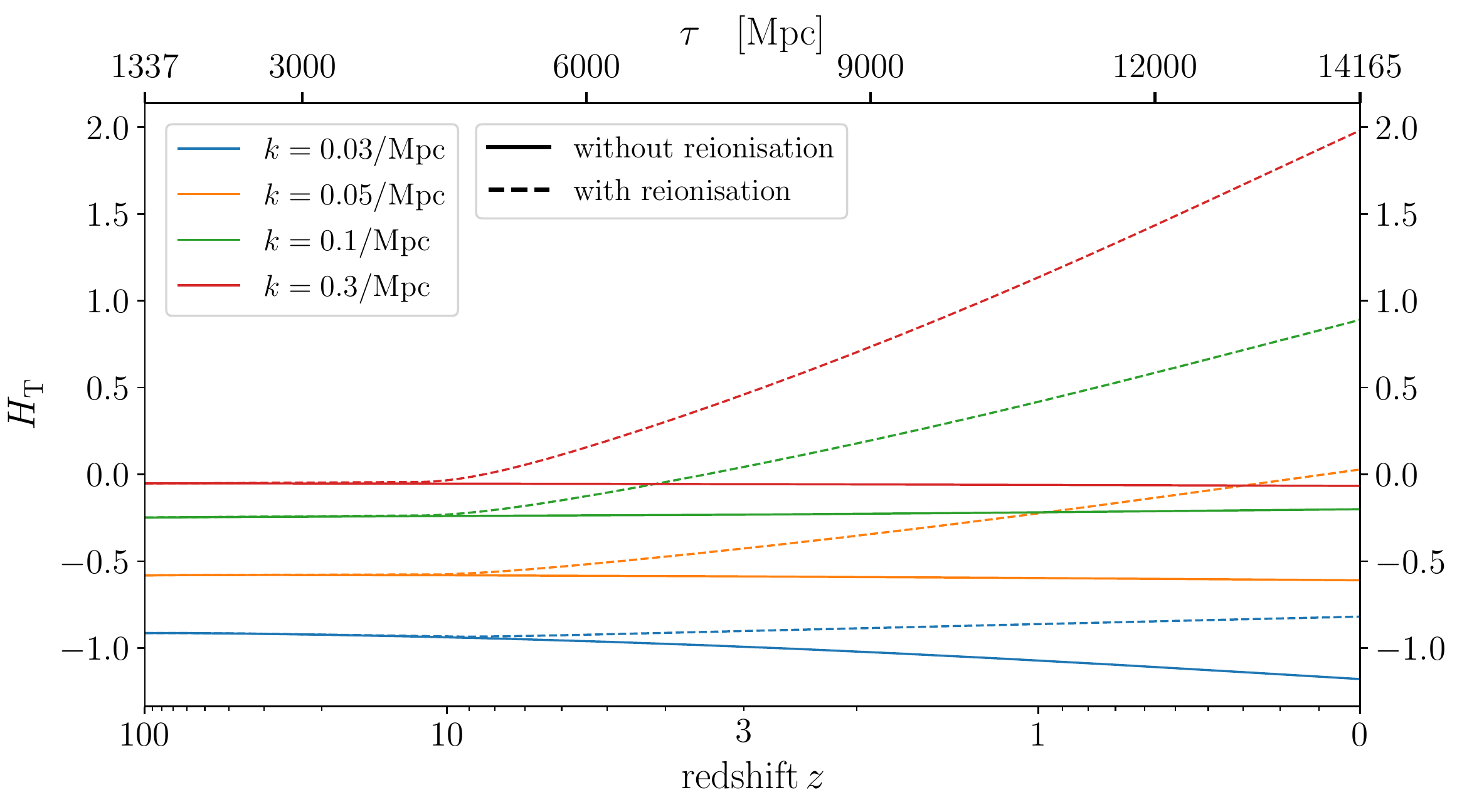} 
\caption{
In solid lines we have plotted the time evolution of $\HT$ for different scales. The dashed lines show the evolution of $\HT$ when taking into account the impact of baryons scattering off CMB photons during reionisation.}
\label{Reionisation Plot}
\end{figure}

In figure~\ref{Reionisation Plot} we can see the impact of reionisation compared to a run where reionisation has been turned off consistently in the thermodynamic computations in \CLASS~and in the differential equation for our gauge transformation. We can see the effect of the Thomson scattering around the reionisation redshift of $z \approx 11$. Afterwards, $\HT$ evolves almost source-free and stays small, so that reionisation can easily be described in our framework.
While this correction remains very small and can hardly be detected in the total matter power spectrum, reionisation does leave a small imprint in the relative distribution between baryons and cold dark matter, as seen in figure \ref{Baryon Bias Plot}.

\subsection{CDM and WDM}\label{sec:CDMWDM}

The previous cases were simple and could be reduced on the level of the metric to a single fluid Nm or N-boisson gauge [cf. section~\ref{sec:Nboisson}], without the need to incorporate a mass modulation. 
As a first example of a more complex model we study here the case of warm dark matter, possibly mixed with cold dark matter, where the mass modulation will be non-zero.

\begin{figure}
\includegraphics[width=\textwidth]{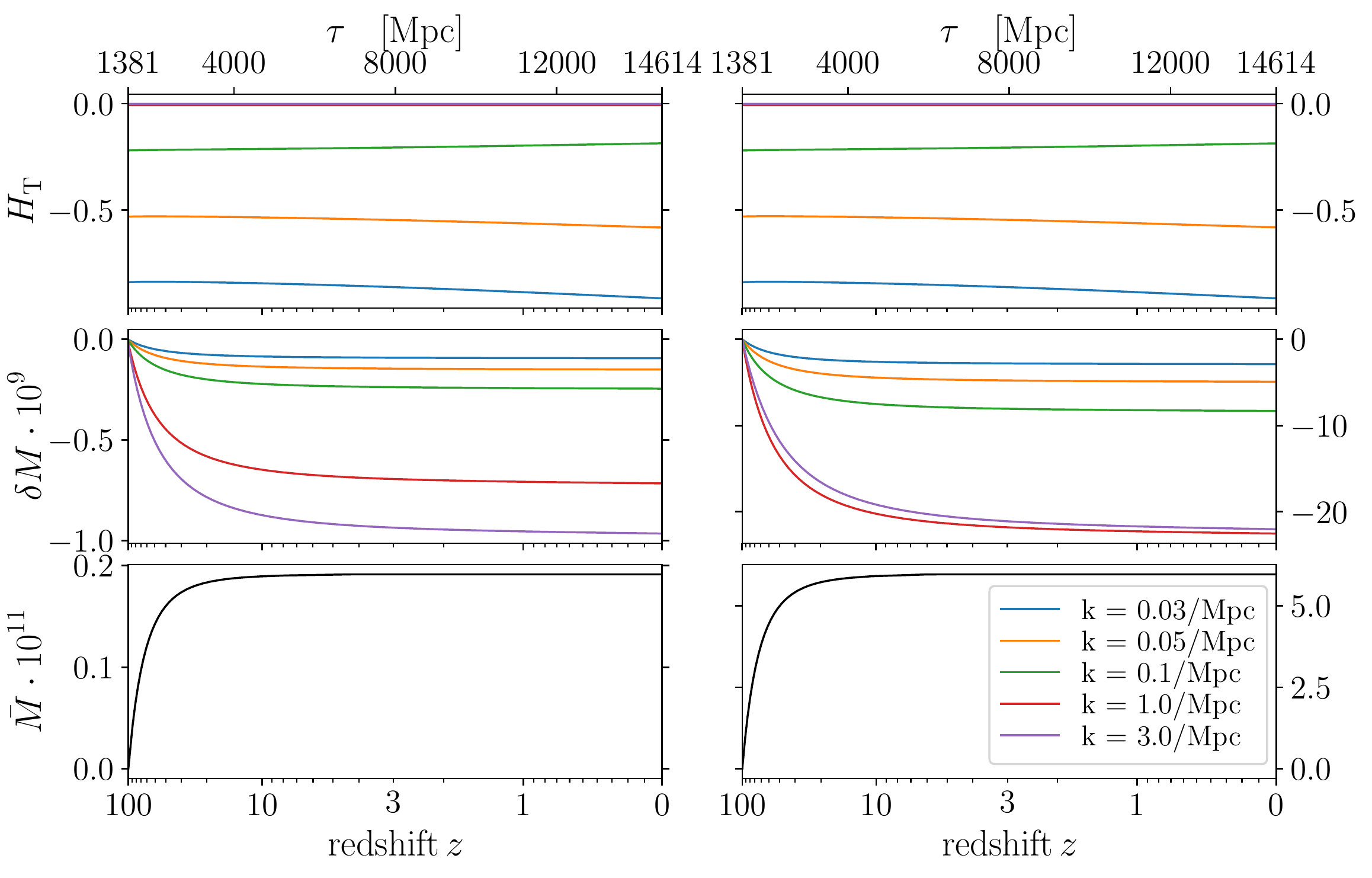} 
\caption{$\HT$, $\mybar{M}$ and $\delta M$, on the left hand side for one WDM species with $m = 5000$ eV making up all of the dark matter, and on the right hand side for one WDM species with $m = 200$ eV making up $5$\% of the total dark matter.}
\label{WDM Plot}
\end{figure}

We have chosen two typical cosmologies with WDM components, one with all of the dark matter being warm and one with $5$\% WDM. We have plotted $\HT$, $\bar{M}$ and $\delta M$ for a simulation initialised at $z=100$ for both cosmologies in figure~\ref{WDM Plot}.

In the case of $100$\% WDM the mass has to be very high in order not to get into conflict with current constraints. This means that the non-relativistic transition of the fluid occurs very early. The mass modulation is non-zero, but smaller than any resolution achievable with state-of-the-art simulations. The evolution of $H_T$ is not distinguishable from the pure CDM case.  For one WDM species with $m = 200$ eV making up $5$\% of the total dark matter the mass modulation is approximately one order of magnitude higher, but still negligible. In these simple cases, WDM can easily be modelled by Newtonian motion gauges. The post-processing is exactly the same as in a CDM-only simulation, all of the differences are already encoded in the initial conditions, which have to be generated from a matter power spectrum in the respective cosmology.

\paragraph{Changing from CDM to WDM via post-processing.}

\begin{figure}
\begin{center}
\includegraphics[width=0.6\textwidth]{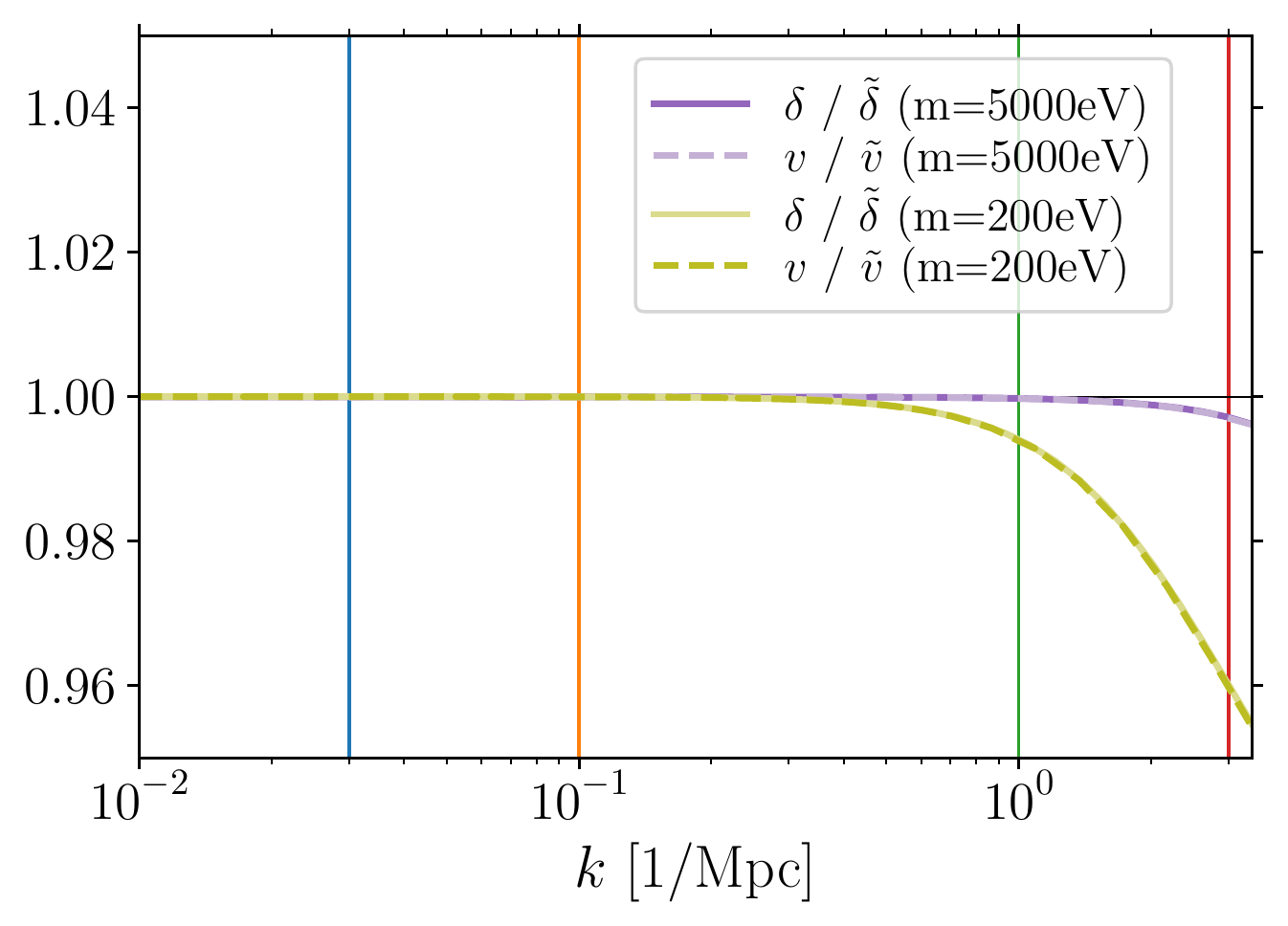} 
\end{center}
\caption{The density and velocity transfer functions that are used to initialise warm dark matter simulations in comparison to a cold dark matter cosmology. The two warm dark matter cases initially have a reduced power on the small scales. The four vertical lines correspond to the $k$-values shown in the following plot.}
\label{WDM Plot CDM init starting}
\includegraphics[width=\textwidth]{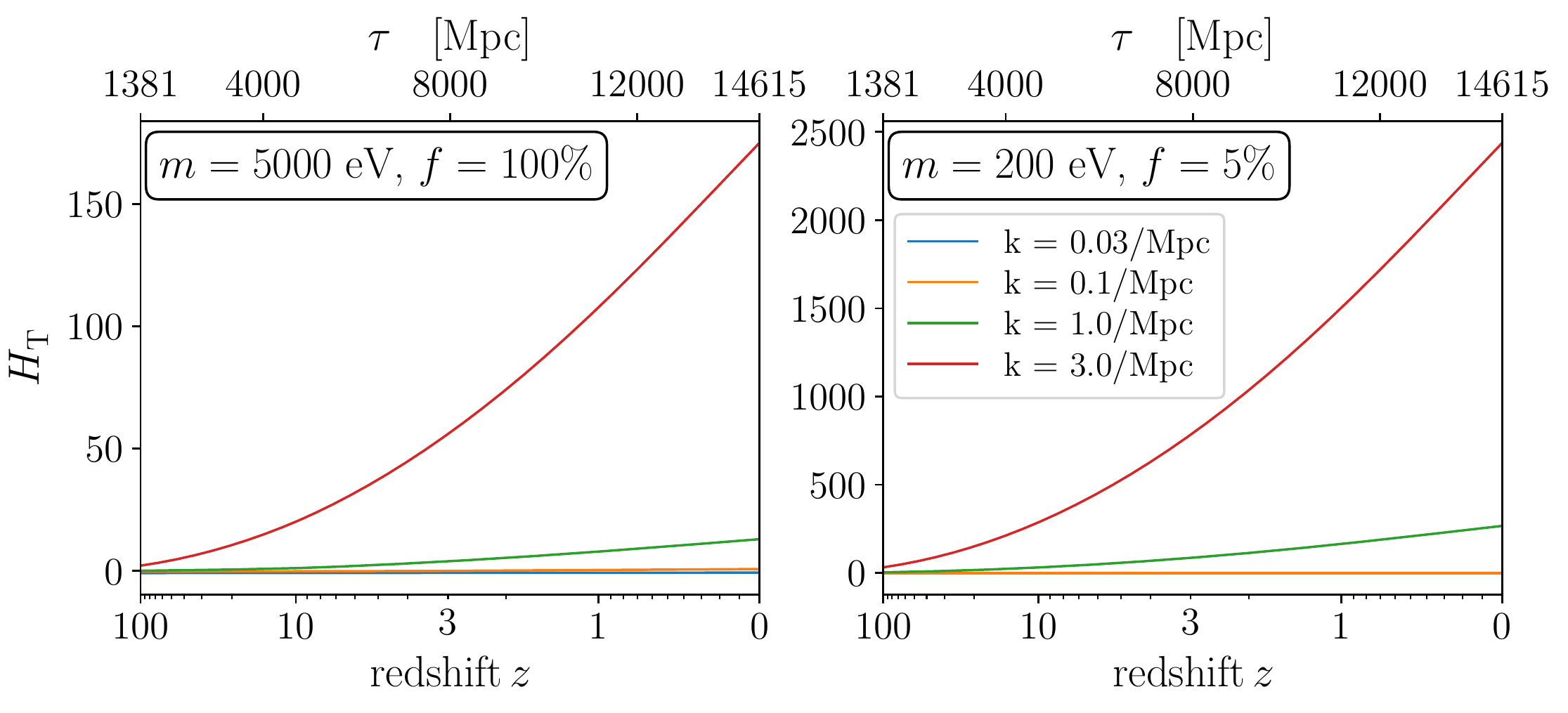} 
\caption{$\HT$ with the CDM initial conditions for both WDM cosmologies discussed in this chapter. In both cases we see that the mismatch in the initial conditions on the small scales (see plot above) requires an $\HT$ starting away from $3\zeta$ that grows significantly throughout the evolution and may become large enough to violate the weak-field assumptions. }
\label{WDM Plot CDM init}
\end{figure}

Since the values of $\HT$, $\bar M$ and $\delta M$ remain very small, one might wonder if we could simply use a CDM simulation with the CDM initial conditions to simulate the effect of warm dark matter. But what is not captured in the above plots are the different initial conditions that one gets in the WDM case, as opposed to CDM case, shown in figure \ref{WDM Plot CDM init starting}. We might further seek to also absorb these into the definition of our Newtonian motion gauge. To do so we would employ the residual gauge freedom to find an Nm gauge where the initial Newtonian densities mimic the ones in a pure cold dark matter cosmology.  

In Appendix \ref{sec:initial conditions} we discuss such initial conditions for $\HT$ and how to calculate them to match the initial conditions of an N-body simulation.
The initial conditions of $\HT$ and $M$ have to be adjusted and we find 
\be
\delta \rho_N \approx \delta \rho_m + 3 \HL \bar{\rho}_m = \delta \rho_m + 3 \bar{\rho}_m (\Phi - \frac{1}{3} \HT) = \tilde{\delta \rho_m} + 3 \tilde{\bar{\rho}}_{m} (\tilde{\Phi} - \tilde{\zeta}) \, ,
\ee
where the tilde denotes quantities from the CDM cosmology. We link the different cosmologies and obtain the initial condition for $\HT$:
\be
\HT = \frac{\delta \rho_m - \tilde{\delta \rho}_m}{\bar{\rho}_m} - 3 \frac{\tilde{\bar{\rho}}_m}{\bar{\rho}_m} (\tilde{\Phi} - \tilde{\zeta}) + 3 \Phi \,.
\ee
The initial derivative of $\HT$ is computed from 
\be
v_N = v_m + \kop^{-1}\dot{H}_{\rm T} = \tilde{v}_m + 3 \kop^{-1} \dot{\tilde{\zeta}} \,,
\ee
which implies that
\be
\dot{H}_{\rm T} = \kop (\tilde{v}_m - v_m) + 3 \dot{\tilde{\zeta}} \,.
\ee
We apply this approach to the two WDM cosmologies we have discussed before, and illustrate the results in figure~\ref{WDM Plot CDM init}. On small scales, where WDM suppresses structure formation, starting from the ordinary cold dark matter initial conditions requires a rather large value of $\HT$. The reason is that the CDM simulation does not have a suppression of the small scales and this now needs to be implemented at the level of the coordinate system. In the case of $m=200$ eV, on small scales $\HT$ quickly reaches values where perturbation theory cannot be trusted anymore. This shows that $\HT$ cannot be used to change the cosmology this drastically. The $m=5$ keV case looks more promising, but care has to be taken because $\HT$ is growing quickly towards the end of the simulation. Its changing values have to be taken into account when analysing multiple snapshots at different redshifts.
We find that CDM initial conditions are generally problematic for the analysis of warm dark matter models. It is often not possible to reinterpret a WDM cosmology based on a CDM simulation plus a linear rescaling of the coordinate system to weak-field precision.
Instead when using appropriate warm dark matter initial conditions $\HT$ remains small as the changes are already contained in the initial conditions and evolve non-linearly from there.

\subsection{CDM and massive neutrinos}\label{sec:nuCDM}

Massive neutrinos can be added in the same way as warm dark matter with the crucial difference that the transition to a massive species happens significantly later, typically during runtime of the N-body simulation. As a consequence we do not only change the initial conditions, but also have a significant evolution of $\bar M$ and $\delta M$ during the later stages. While warm dark matter can be well described in an ordinary single fluid approach from modified initial conditions, the neutrino case is much more complex. In the following sections we briefly summarise the differences between neutrinos and warm dark matter from the perspective of Nm gauges. 

\paragraph{Neutrino pressure.}
The pressure of massive neutrinos decays over the evolution of the Universe and sharply falls during the non-relativistic transition. 
The pressure perturbations and the anisotropic stress, however, do not decay that quickly. After the transition we are dealing with a massive fluid that still has a very complex phase space. In the Euler equation we can see that the impact of the pressure perturbation is even enhanced on the small scales by a factor of $k$. This term, while still being small, is potentially problematic for our method as even a small deviation from the Newtonian dynamics on the small scales must be compensated by a significantly larger correction in $\HT$, because the impact of $\HT$ on the motion of massive particles is suppressed on the small scales. Correcting a small deviation may therefore involve large metric perturbations that may be incompatible with the weak-field assumption.   

\begin{figure}
\includegraphics[width=\textwidth]{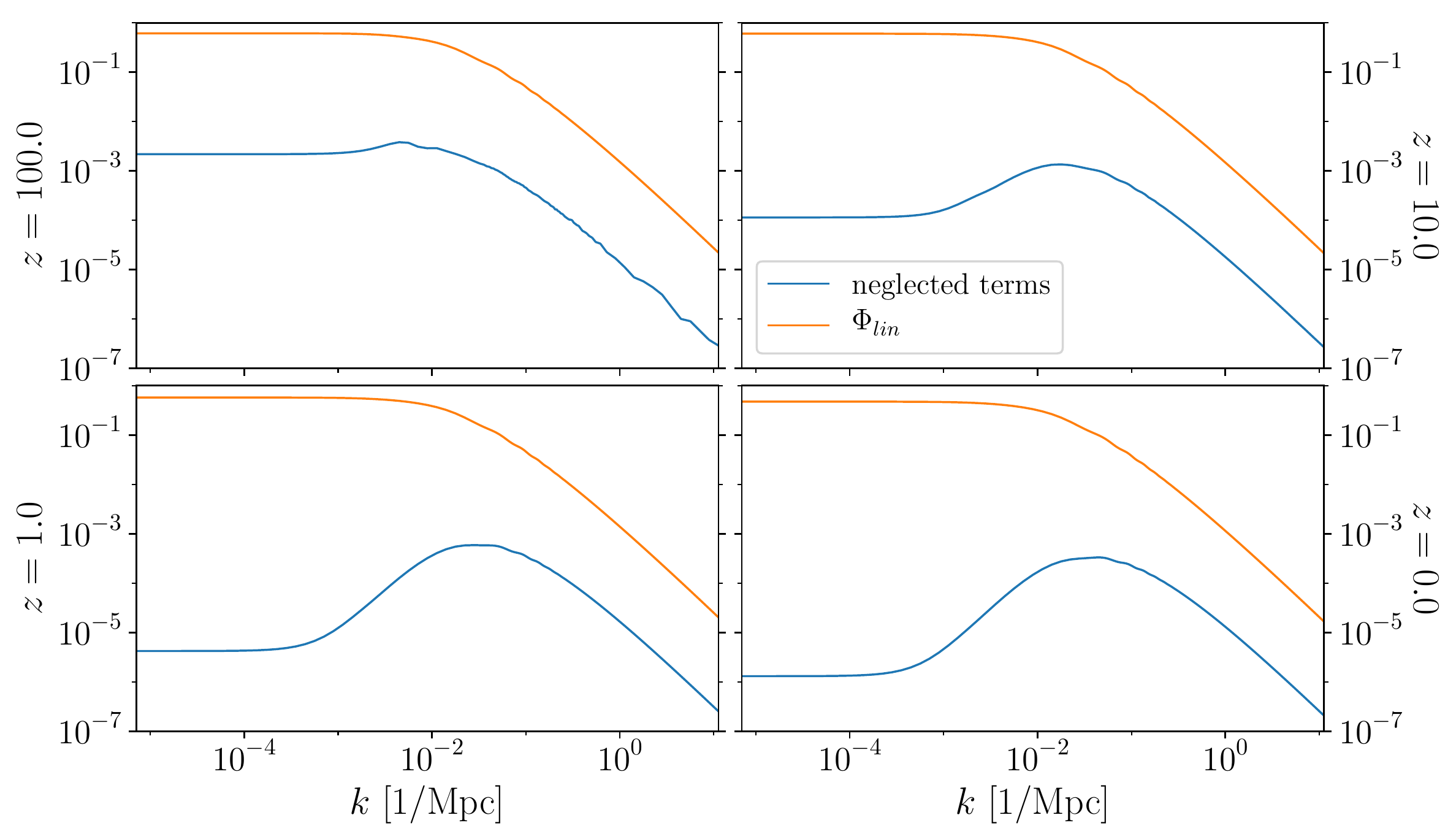} 
\caption{The neglected terms in the Euler equation proportional to neutrino pressure and shear perturbations compared to the linear gravitational potential for $\sum \limits_{i} m_{\nu_i} = 170$meV.}
\label{neglected terms}
\end{figure}

We compare the size of these contributions to the other sources in the massive Euler equation and find that at all times and on all scales the impact of pressure perturbations is suppressed by at least two orders of magnitude compared to the linear gravitational potential. In addition, there are several other factors that reduce the overall importance of these terms. At early times when the pressures are largest these are considered as a part of the relativistic fluid and consistently included in the Boltzmann code. At the later times and small scales, the linear potential receives significant non-linear corrections and the relative impact of the pressure terms is also suppressed in comparison. Finally in the intermediate region, the pressure terms do not provide a consistent force that is integrated along the particle trajectories, but largely cancels out as the particles propagate. This is opposed to the gravitational potential that over a long time builds up the matter velocities until they eventually become large and non-perturbative. All other terms which are influenced by the presence of a neutrino pressure and anisotropic stress, such as the separation of the potentials are taken into account at all times. For these reason we neglect the terms in question and this approximation is of comparable accuracy as the weak-field limit. 
We plot the pressure and anisotropic stresses as they appear in the shared fluid Euler equation and the linear potential for comparison in figure \ref{neglected terms}. Note that this represents a worst case scenario since we plot only the linear potential and do not take into account that the terms are consistently included in the relativistic species before the neutrino transition.

Note that ref.~\cite{Dakin:2017idt} found a significant effect on the neutrino powerspectrum from replacing the pressure perturbation by the density perturbation according to the equation of state. But that is not the approximation we are making here. We are \emph{only} removing the impact of the non-relativistic neutrinos in the shared fluid Euler equation for the massive species. The pressure of the relativistic part of the neutrino perturbations is always kept within the Boltzmann-Einstein solver and its influence on the matter evolution is included.

\paragraph{Initial conditions for neutrino simulations.}

There are two distinct methods to initialise simulations in the Newtonian motion framework, explained in detail in \cite{Fidler:2016tir}. These are related to the residual gauge freedom on the boundary, allowing the choice of arbitrary values for $\HT$ and $\dot{H_{\rm T}}$. In the multifluid case we may additionally choose any value for $\bar{M}$ and $\delta M$.

The initial conditions can be derived directly from the relativistic perturbations at the initial time in a simple gauge choice, typically the N-boisson gauge [cf. section~\ref{sec:Nboisson}]. This means that we fix the residual boundary conditions at the time when the simulation is initialised. 
We call this approach the {\it forwards method.}

Alternatively, we may fix the residual gauge freedom at the final time. The simulation then does not start from a simple gauge, but it ends with the one. In order to find such a gauge, the equations for the metric perturbations have to be solved "backwards", thus we call it the {\it backwards method}. The corresponding initial conditions describe how a Universe following Newtonian equations of motion would look initially, given that it resembles our relativistic Universe today. 
 
In the forwards method, $\bar{M}$ and $\delta M$ are initially zero and the initial conditions represent all the massive species present in the Universe at that time. However, the mass modulation rises once the neutrinos become non-relativistic. For the remaining part of the simulation it does not include all matter and the metric perturbation $\HT$ is required to absorb this discrepancy. As a consequence $\HT$ is continuously growing and may become very large, challenging the weak-field assumptions. 

The alternative is the backwards method. In this case the mass modulation is vanishing at the final time and is only different from zero before the neutrino transition. In order to correctly represent the neutrino perturbations in the Newtonian simulation at the late times, the corresponding initial conditions must contain a decaying mode, designed such that it mimics the neutrino perturbations around decoupling. Initially the simulation deviates significantly from the forwards method, but after the neutrino transition we are left with a massive fluid that now has the correct mass and the correct neutrino perturbations imprinted in it. From that point onwards the simulation can be interpreted using a simple N-boisson gauge. 

For cold dark matter this backwards method is equivalent to the commonly applied method of backscaling, where the present day linear power spectrum is scaled back to the initial time using the linear growth function. Note that this does no longer hold in the case of massive neutrinos since we require and additional decaying mode that would not be captured by backscaling. Instead the initial conditions must be derived from within the Newtonian motion framework after fixing the residual gauge freedom, and not the power spectrum, at the final time.

In order to find this solution we employ a simple root finding algorithm that tests various combinations of $\HT$, $\dot{H}_{\rm T}$ and $\delta M$ until it finds a solution with suitable features. This method can then be used to compute optimised initial conditions for N-body simulations including the impact of neutrinos. 

\paragraph{Results.}

\begin{figure}
\includegraphics[width=\textwidth]{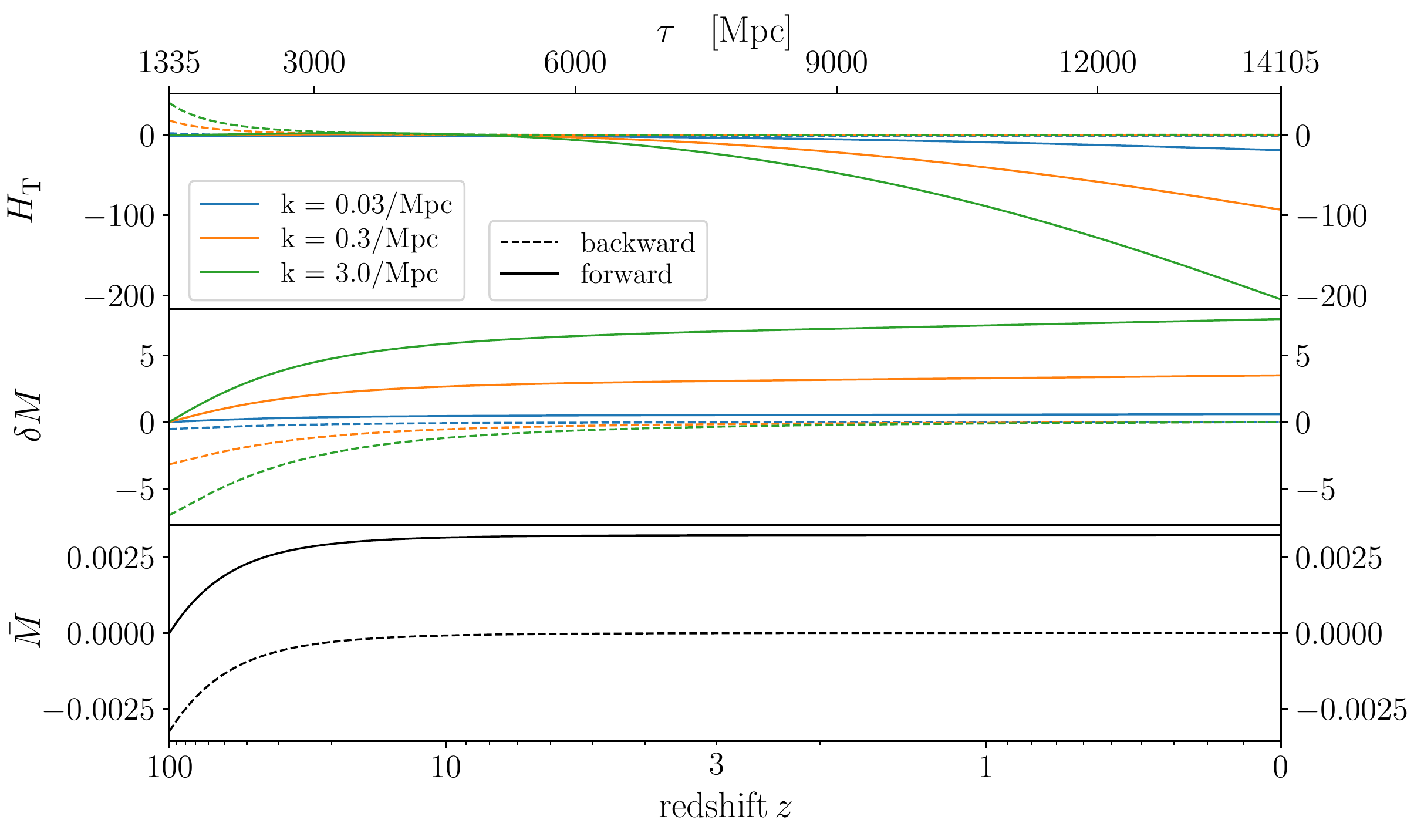} 
\caption{Comparison of the forwards and backwards approach for a combined neutrino mass of 170meV. In the latter case the simulation is started from modified initial conditions, but the evolution of these then quickly approaches the correct late Universe cosmology. The metric potential $\HT$ evolves until the neutrino transition and then stays on $\HT = 3\zeta$. In the forwards method we find a constant mass modulation at the later times that causes a continuous growth of $\HT$. Overall the value of $\HT$ is significantly smaller in the backwards analysis and remains in agreement with the weak-field assumptions at all times.}
\label{170 Neutrino forward}
\end{figure}

\begin{figure}
\includegraphics[width=\textwidth]{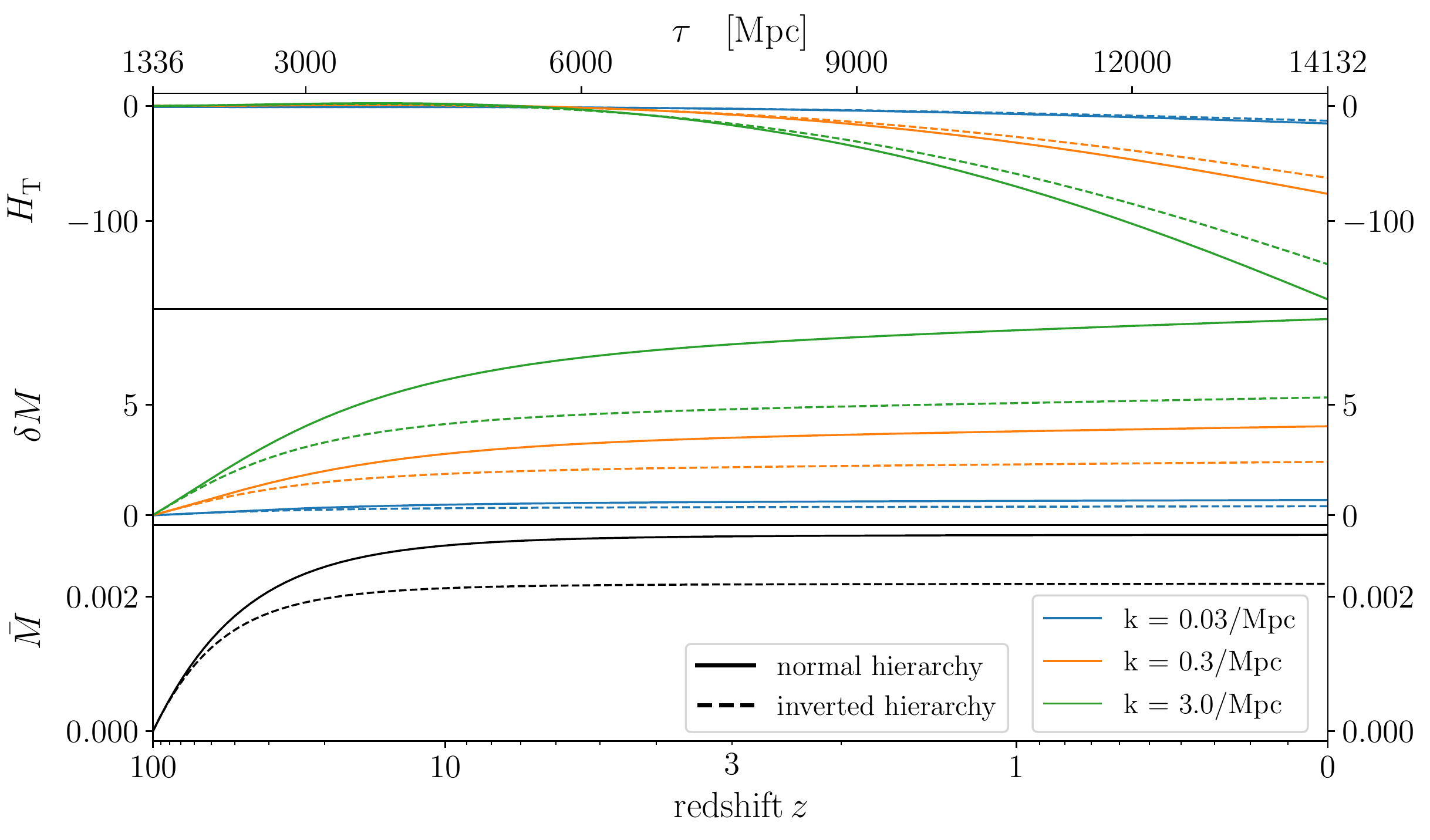} 
\caption{A forward neutrino Nm gauge for a combined neutrino mass of $99$\,meV for normal and inverted mass hierarchy. The size of the metric potentials remain comparable but the temporal evolution is changed as the neutrinos become non-relativistic at different times. The mass modulations stay small throughout the entire evolution while, $\HT$ grows on the small scales and at late times. On the analysed scales the weak-field assumptions remain valid.}
\label{hierarchy comparison}
\end{figure}

We show the forward and backward solutions for a combined neutrino mass of $170$meV in figure~\ref{170 Neutrino forward} as well as a comparison of the normal and inverted hierarchy for $99$\,meV in figure~\ref{hierarchy comparison}. We have chosen $99$\,meV to correspond to the minimum allowed mass for an inverted hierarchy maximising the relative difference between the two cases. The larger mass of $170$meV was chosen to represent the most problematic mass range for the Nm gauge approach. For smaller masses neutrinos have an overall smaller impact and $\HT$ can remain small. But for larger masses the transition happens earlier and a larger fraction of the neutrinos is directly absorbed in the initial conditions for the massive species. For very large masses this eventually produces the warm dark matter case where only the initial conditions are modified and $\HT$ can remain constant. 

In the forward case the mass modulation rises during the non-relativistic transition before flattening out at a value of two permille, which is of the expected magnitude. The perturbation $\delta M$ evolves similar to $\bar M$ and remains perturbative on all scales. Due to these sources $\HT$ continuously grows and and reaches values in the order of a few hundreds that slowly approach the non-perturbative regime. 

Using the backwards method the mass modulation starts out negative (i.e. too much mass in the simulation) but approaches zero quickly after the non-relativistic transition. The metric perturbation starts away from the equilibrium value of $\HT= 3\zeta$, but quickly converges to it after the non-relativistic transition. Afterwards the metric becomes trivial. In this case all perturbations stay small and the weak-field limit can safely be applied on all scales. In addition the simulation may be analysed on the much simpler N-boisson metric compared to the evolving forwards method. 

Generally speaking the backwards case is well suited to simulate massive neutrinos, while the forwards method remains valid for sufficiently small neutrino masses. 

We also compare the difference between a normal and inverted hierarchy for the minimal possible mass of $99$\,meV in the forwards method. We find that the metric remains perturbative while only small differences between the two cases are visible. For larger masses these are even smaller.   

\begin{figure}
\includegraphics[width=\textwidth]{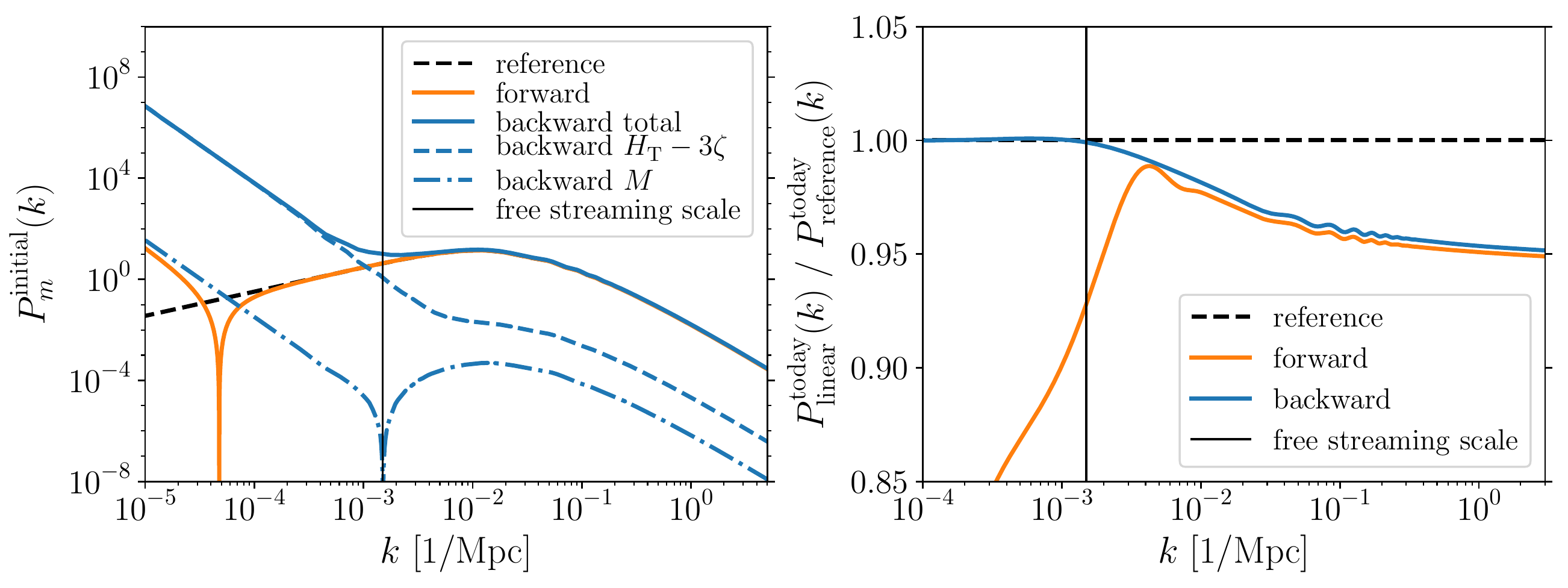} 
\caption{Matter power spectra for the generation of initial conditions in numerical simulations, compared to a massless neutrino reference cosmology. The left plot refers to the initial time ($z=100$) while the right plot is the linear present day power spectrum of $\delta_N$, resulting from those initial conditions. In the forward case, we find that the initial conditions remain close to the reference cosmology. At the present time we find a difference on all scales, but it has to be interpreted on the non-trivial metric of a Nm gauge. In the backwards case we find significantly changed initial conditions, including decaying modes, while the results can be interpreted on the much simpler N-boisson gauge at the present time. At that time we see that the large scale power remains identical to the reference cosmology, while neutrino streaming suppresses the small scale structures. }
\label{Neutrino initial conditions}
\end{figure}

In figure \ref{Neutrino initial conditions} we summarise the power spectra that can be used to generate the initial conditions for Newtonian simulation for a combined neutrino mass of $74 \text{meV}$, where both the backwards and forwards approach are valid. These are the power spectrum of the Newtonian over-density 
\be
\delta_N = (1-\bar{M})\delta_m + 3 \HL - \delta M \,,
\label{eq:back-ini}
\ee 
where $\delta_m = \sum \limits_X (\rho-3p)_X\delta_X$ is the sum over all species that contribute to the massive species. At the initial time these are mostly baryons and CDM, but there is a small non-vanishing contribution from the neutrinos. The initial velocity cannot be computed from the density under the assumption of a pure growing mode. Instead it is set independently by $v_N = v_m$, where $v_m$ is the sum over the massive velocities. Note that the density $\delta_m$ only depends on the temporal gauge choice and is identical for all Newtonian motion gauges. The change in the initial conditions is introduced from the changing values of $M$, $\delta M$ and $\HL$. On the other hand, the velocity $v_m$ does depend on the spatial gauge choice and varies between the different Newtonian motion gauges. In order to make this explicit we can write the initial Newtonian velocity in reference to the N-boisson gauge velocity
\be
v_N = v^{\rm N-boisson}_m + \kop^{-1}(\dot{H}_{\rm T} - 3\dot{\zeta}) \,.
\ee
Note that the N-boisson gauge density is, to linear order, the Poisson gauge density, and the N-boisson gauge velocity is almost identical to the Poisson gauge velocity since $\dot{\zeta} \approx 0$. 

In the left panel of figure~\ref{Neutrino initial conditions} we present the initial power spectrum of $\delta_N$ used to set up a Newtonian N-body simulation.
We compare to a reference cosmology with massless neutrinos, that is initialised on the N-boisson gauge. In this case the Newtonian density is 
\be
\delta_N^{\rm ref} = \delta_m + 3 (\Phi- \zeta) \,,
\label{eq:ref-cos}
\ee 
which is identical to the total matter (or approximately the synchronous) gauge density. The forward case is also initialised on the N-boisson gauge, but now $\delta_m$ does include a small correction from the massive neutrinos. In the reference cosmology the Poisson gauge large-scale divergence of the density is canceled by the Bardeen potential $\Phi$, leading to a finite Newtonian density on all scales. But this cancellation does not hold for  relativistic species such as the neutrinos. As a consequence the otherwise small neutrino correction becomes dominant on the largest scales.

The backwards analysis is initialised on a very complex metric, but quickly approaches the N-boisson gauge after the neutrino transition. We find that the initial conditions in this case differ significantly from the forward case. In addition to the full matter power spectrum, we show the auto-powerspectra of the various additional contributions from Eq.\,\eqref{eq:back-ini}. Contributions due to the mass modulation (labelled $M$ in the plot) always play a subdominant role, while the deviation of $\HT$ from the N-boisson gauge (labelled $\HT - 3 \zeta$) is dominant on the large scales and about three orders of magnitude below the combined power-spectrum on the smaller scales. The plot does not show the cross-terms that are responsible for a change of the total power on the small scales of a few percent. 

The right panel shows the present day matter power spectrum that would be obtained in the Newtonian simulation from these initial conditions, compared to the reference cosmology. Here we have only used linear perturbation theory, but an N-body simulation would compute this evolution in full non-linearity. This implies that we have not taken into account any mode-coupling and the output can be related on a one-to-one basis to the initial value at the same wavevector $k$.

The backwards case is constructed such that we end up on the N-boisson gauge, and thus we find the expected impact of massive Neutrinos on the power spectrum. On the small scales that is a relative suppression due to a change in the background cosmology and the few percent change in the initial power spectrum. On the larger scales the initial conditions are such that the Newtonian simulation recovers the reference cosmology. Note that the large values of the initial power spectrum are partially compensated by the initial velocity that together form a decaying mode imprinting the correct neutrino perturbations at the transition.   

The forwards case has a simple metric initially at the price of having a complex Nm gauge at the present time. In this case the initial conditions remain much closer to the reference cosmology apart from the impact of the neutrinos on the largest scales. Consequently, at the final time, the power on these scales is significantly reduced. However, we show the Newtonian density that still needs to be interpreted on the complex Newtonian motion gauge metric. When doing this, we recover the same result as in the backwards case for the relativistic density. 

It is further interesting to look at the difference between the forward and backward case at the small scales. In the forward case the power is smaller, which is corrected by the non-vanishing mass modulation when constructing the relativistic densities. In the backwards case this is not necessary as the effect is already imprinted in the modification of the initial power spectrum. Running a neutrino simulation without modifying the initial conditions, or using a relativistic interpretation of the results, would overestimate the suppression by this amount. This mismatch is of percent-level, but grows quickly with the neutrino mass.  

In order to gain accurate results, this analysis is of course not sufficient. A non-linear Newtonian simulation is required to study the evolution from these initial conditions accurately, including significant corrections on small scales.

\paragraph{Interpretation of the simulation.}

For a given snapshot first the mass modulation is incorporated by changing the particles masses. 

The relativistic density is $\delta\rho = \delta\rho_N + 3\rho\HL - \rho M$. We want to express it instead as $\delta\rho = \tilde{\delta\rho}_N + 3\rho \HL$ with a new density that is not subject to a mass modulation. 
The densities are $\rho_N = \bar\rho + \delta\rho_N$ and $\tilde{\rho}_N = \bar\rho + \tilde{\delta\rho}_N = \bar\rho + \delta\rho_N -\rho M $.
Therefore the ratio of densities evaluates to
\be
\frac{\tilde{\rho}_N }{\rho_N}  = 1 - M \,,
\ee
and this is the ratio by which the masses should be changed. 
After this step the particles may be spatially displaced to a gauge with $\HT = 3\zeta$. 
We have now taken into account the relativistic effects and obtain the output in the N-boisson gauge. For a discussion how to resolve the individual fluids we refer to Appendix \ref{sec:resolve}.

\section{Conclusions and future work} \label{sec:concl}

We generalise the concept of Newtonian motion gauges to systems with multiple species, allowing these species to transit from relativistic to non-relativistic 
during the time of the simulation. This allows us to apply the Newtonian motion framework to scenarios including warm dark matter and massive neutrinos in addition to ordinary cold dark matter. We focus only on scalar modes for simplicity, but vectors and tensors contribute in the same way as in the single fluid case \cite{NLNM}.
At the leading order in the weak-field approximation, they have no impact on the evolution of matter and they can be reconstructed from the output of the Newtonian simulation and the Einstein-Boltzmann code.

We confirm that the usually employed method of initialising baryons and CDM as a single fluid is in agreement with general relativity in the weak-field limit. The only approximation is neglecting the mixing stress and pressure which can be included in the simulation by adjusting the initial conditions. We provide a simple method for assigning species to the simulated particles, but find that the ratio between baryon and CDM densities remains close to the background value. In addition we study the impact of reionisation on the evolution and find a small effect due to the friction added to the baryons. This is mainly visible in the baryon-to-CDM ratio, but it remains well below experimental sensitivity even there.

Warm dark matter is sufficiently non-relativistic by the time when simulations are initialised. As a consequence the main modification is the impact on the initial conditions for the simulation. Afterwards there are only tiny corrections to the subsequent motion of the particles from the remaining temperature, so ordinary Newtonian N-body simulations are well suited to simulate warm dark matter cosmologies. The final results, combining changes in the initial conditions and to a much lesser degree in the evolution, then differ significantly from cold dark matter cases since the discrepancies in the initial matter distribution are non-linearly evolved until the present time. 

Massive neutrinos provide a more interesting application of the Newtonian motion gauge framework. They are initially hot but cool down during the simulation. We find that a mass modulation and a non-trivial coordinate system are required to keep the centre-of-mass evolution Newtonian. These modifications are relevant on all scales. In addition to the forwards approach without major changes to the usual N-body initial conditions we discuss a backwards method using optimised initial conditions that ensure that 
the simulations can be interpreted on the simple N-boisson gauge at late times. 
These differ from the commonly employed backscaling initial conditions by including a tailor-made decaying mode that imprints the desired neutrino perturbations at the time of neutrino decoupling. Our method also improves the convergence within the weak-field limit by utilising that a small change in the initial condition may cause a much bigger, non-linear modification of the final results. 

We find that a Newtonian analysis of a simulation with massive neutrinos is not only inconsistent on the larger scales, but also causes percent-level corrections on the small scales that may be enhanced by the non-linear evolution. Our method in principle allows simulations including neutrinos and general relativity at almost the same computational cost of ordinary CDM-only N-body simulations. This is possible since the non-trivial phase space of the neutrinos is kept entirely within a linear-Boltzmann code 
and the centre-of-mass motion is obtained by Newtonian N-body simulations.

Relativistic simulations including neutrinos are becoming an important challenge for the analysis of future large scale missions such as EUCLID. The NM gauges have already been demonstrated to work in the massless neutrino case, from the very large super-horizon to the galactic scales.
In the present paper we generalise this method to consistently include neutrinos in weak field relativity. However, our method is not directly compatible with Newtonian N-body simulations, but relies on starting these from modified initial conditions. Without running such a simulation there is no way to compare to existing methods and explicitly demonstrate that we can accurately describes the impact of non-linear neutrinos on the small scales. Instead for now we only provide theoretical arguments and demonstrate that our method does work on the linear scales. We are working on a proof of concept using a modified N-body simulation, but this is left for future work while in the present paper we only introduce the theoretical framework. 

\section*{Acknowledgements}

We thank Cyril Pitrou for support in the use of XPAND, and Daniele Bertacca, Julian Adamek, Jacob Brandbyge, Steen Hannestad for useful discussions. 
The work of CR is supported by the DFG through the Transregional Research Center TRR33 ``The Dark Universe''.   
The work of KK has received funding from the STFC grant ST/N000668/1, as well as the European Research Council (ERC) under the European Union's Horizon 2020 research and innovation programme (grant agreement 646702 ``CosTesGrav").

\appendix

\section{Separating the components of the combined fluid}\label{sec:resolve}
In the case of CDM and photons the output of the N-body simulation may directly be understood as the dark matter component while the Einstein--Boltzmann code provides the photons. 
But a multifluid simulation only computes $\rho_m$, the center-of-mass density. Disentangling this into the individual components is a secondary task that can be done as a part of post-processing.

The best option is to identify a combination of the massive species that is shielded from the non-linear collapse. This combination may then be evaluated to linear order in a Einstein--Boltzmann code and can be used to break the shared fluid back into the individual ones. This approach may be applied in the baryon plus CDM case and we define the difference of the overdensities
\be
\delta_{\rm cdm} - \delta_{b} = \delta_{\Delta}.
\ee
We further define the relative velocity $v_{\Delta} = v_{\rm cdm} - v_{\rm b}$, which allows us to replace
\be
v_{\rm cdm} = v_m + \frac{\rho_{\rm b}}{\rho}v_\Delta\,, \;\;\;\; v_{\rm b} = v_m - \frac{\rho_{\rm cdm}}{\rho}v_\Delta\,.
\ee
We employ the relativistic continuity equations to find 
\be
	\dot{\delta}_{\Delta} = \hat\nab^i v_{\rm cdm}\hat\nab_i \kop \delta_{\rm cdm} - (1 +\delta_{\rm cdm})\kop v_{
	\rm cdm} - \hat\nab^i v_{\rm b}\hat\nab_i \kop \delta_{\rm b} + (1 +\delta_{\rm b})\kop v_{
	\rm b}.
\ee
In the relativistic Euler equation for the relative velocity the gravitational source cancels and therefore it decays with the expansion of the Universe. It is consistent to postulate that $v_\Delta = \mathcal{O}(\epsilon)$.
We then obtain to weak-field order
\be
(\partial_\tau + \Hc)\hat\nab^i v_\Delta = \hat\nab^j v \hat\nab_j \kop \hat\nab^i v_\Delta + \hat\nab^j v_\Delta \hat\nab_j \kop \hat\nab^i v\,.
\ee
It is clear that this will generate a density difference of order $\kappa\epsilon$ and to leading order we have
\begin{align}
	\dot{\delta}_{\Delta} =&  \hat\nab^i v_\Delta\hat\nab_i \kop \delta - (1 +\delta)\kop v_\Delta + \hat\nab^i v \hat\nab_i \kop \delta_\Delta   - \delta_\Delta \kop v\,.
\end{align}
While the relative velocity and density are not enhanced on the small scales, the leading order equations are non-linear and cannot be solved in a linear Einstein--Boltzmann code.
For the relative velocity these non-linear terms are diffusion damping and a term suppressing the relative velocity in collapsing regions. The linear approximation is therefore only overestimating the relative velocity. Since the linear solution for the source-free relative velocity is already small we conclude that linear perturbation theory is sufficiently accurate.  

Non-linear corrections are more important for the relative density. 
Instead we define the local concentration coefficient:
\be
\sigma = \frac{\delta_\Delta}{1+\delta^N}\,.
\ee
It's time evolution is given by
\begin{eqnarray}
\dot{\sigma} &=& -\kop v_\Delta + \hat\nab^i v \hat\nab_i \kop \sigma + \frac{ \hat\nab^i v_\Delta\hat\nab_i \kop \delta  }{1+\delta^N} \,,
\end{eqnarray} 
and a local collapse does not change the local concentration. The remaining non-linear terms are diffusion damping and a term proportional to the relative velocity, which is negligible when non-linear effects become relevant. Again the linear theory proves to be a rather accurate estimate. 

We may use the concentration to assign particles to be either cold dark matter or baryons. The ratio of cold dark matter to the total matter is
\be
\frac{\bar{\rho}_{\rm cdm}}{\bar\rho} (1 - \frac{\bar{\rho}_b}{\bar\rho}\sigma)\,,
\ee
while the fraction of baryons is 
\be
\frac{\bar{\rho}_{\rm b}}{\bar\rho} (1 + \frac{\bar{\rho}_{\rm cdm}}{\bar\rho}\sigma)\,.
\ee

\paragraph{Resolving neutrinos.}

For neutrinos the situation is more complex since the relative neutrino velocities stay relevant until much later in the simulation. We may employ the techniques presented above, but these will only provide a first guess to the neutrino distribution. 

Instead we can find accurate neutrino distributions by evolving test particles in the numerical simulation: The shared fluid allows us to compute all relevant metric perturbations since these are only sensitive to the shared fluid. This means that we have all the information required to move test particles that could either represent neutrinos or cold dark matter. We thus propose to simulate a passive spectator fluid, that itself is not used to compute the simulation potential, representing cold dark matter in the gravitational field of the shared fluid. This will provide a realisation of the dark matter distribution consistent with general relativity. Combining the shared fluid and the separator representing dark matter, we can infer the neutrino fluid. The cost of this approach roughly doubles the price of the N-body simulation since we not only evolve the particles representing the shared fluid, but also an comparable number of spectator particles.  

\section{Initial conditions for the gauge transformation}\label{sec:initial conditions}
As discussed in \ref{sec:gauge}, the spatial gauge condition emerges as a second order differential equation for $\HT$. It does not require a specific set of initial conditions, so we are left with two remaining degrees of freedom at the time of initialisation of the simulation. Using these, we can calculate a Newtonian motion gauge for any set of initial conditions used in the simulation. At the initial time the linear approximation is very accurate and
linearising equation (\ref{definition of simulation density})
we find:
\be
\delta \rho_N = \delta \rho_m + 3 \HL \rho_m\,.
\ee
An initial non-zero mass modulation should be absorbed into $\rho_N$, so it does not appear here. We can now plug in the Bardeen potential, which evaluates to $\Phi=\HL+\frac{1}{3}\HT$ for our temporal gauge condition. This shows the relation to $\HT$:
\be
\delta \rho_N = \delta \rho_m + 3 \bar{\rho}_m \left(\Phi - \frac{\HT}{3}\right)\,.
\label{delta rho_N}
\ee
The derivative of $\HT$ already appears in the Newtonian velocity:
\be
v_N = v_m + \frac{\dot{H}_{\mathrm{T}}}{k} \, ,
\label{v_N}
\ee
where $v_m$ is evaluated in Poisson gauge, which has the same temporal gauge condition but additionally fixes $\HT^{\text{Poisson}}=0$.
Using these equations we can solve for $\HT$ for any set of initial conditions. 

If a simulation is initialised using density and velocity perturbations from a different cosmology calculated in N-boisson gauge [cf. section~\ref{sec:Nboisson}], this corresponds to
\be
\delta \rho_N = \delta \tilde{\rho_m} + 3 \tilde{\bar{\rho}}_m\left(\tilde{\Phi} - \tilde{\zeta}\right) \, ,
\label{delta rho_N tilde}
\ee
where the tilde denotes quantities from this different cosmology and where we have inserted the N-boisson gauge condition $\tilde{H}_{\mathrm{T}}=3\tilde{\zeta}$. We can then equate \ref{delta rho_N} and \ref{delta rho_N tilde} under the assumption that both cases have the same $\delta^N$ initially:
\be
\delta \rho_m + 3 \bar{\rho}_m \left(\Phi - \HT\right) = \delta \tilde{\rho_m} + 3 \tilde{\bar{\rho}}_m\left(\tilde{\Phi} - \tilde{\zeta}\right)\,.
\ee
Solving for $\HT$ yields:
\be
\HT = \frac{\delta \rho_m - \tilde{\delta \rho}_m}{\bar{\rho}_m} - 3 \frac{\tilde{\bar{\rho}}_m}{\bar{\rho}_m} (\tilde{\Phi} - \tilde{\zeta}) + 3 \Phi\,.
\ee
The corresponding initial velocity perturbation in N-boisson gauge is
\be
v_N = \tilde{v}_m + \frac{3\dot{\tilde{\zeta}}}{k} \, ,
\ee
which we can equate with \ref{v_N} analogously:
\be
v_m + \frac{\dot{H}_{\mathrm{T}}}{k} = \tilde{v}_m + \frac{3\dot{\tilde{\zeta}}}{k}\,.
\ee
Solving for the derivative of $\HT$ we get:
\be
\dot{H}_{\rm T} = k (\tilde{v}_m - v_m) + 3 \dot{\tilde{\zeta}}\,.
\ee

Using these initial conditions, we can use a standard Newtonian simulation and interpret it in a cosmology different from the one it was originally run in. All we need to do is calculate the linear perturbations in both cosmologies at the time when the simulation is initialised, calculate $\HT$, and displace the particles accordingly. Then, we can proceed to do any state-of-the-art analysis of the resulting particle distribution. This method is essentially a linear expansion around the non-linear N-body simulation and has a limited range of validity, that can be explored by looking at the size of the corresponding $\HT$. 
In principle, this could be used to sample the cosmological parameter space with N-body simulations without having to run a new simulation for every parameter point, given that a close enough non-linear simulation was already computed.

\bibliographystyle{JHEP}
\bibliography{references}

\end{document}